\documentclass[12pt,preprint,url]{aastex}
\usepackage{amsbsy}
\usepackage{amsmath}
\usepackage{lineno}
\newcommand{\be}{\begin{equation}}
\newcommand{\ee}{\end{equation}}

\def\ltsima{$\; \buildrel < \over \sim \;$}

\def\lsim{\lower.5ex\hbox{\ltsima}}
\def\gtsima{$\; \buildrel > \over \sim \;$}
\def\gsim{\lower.5ex\hbox{\gtsima}}

\shorttitle{Nuclear pinning and pulsar glitch rates}
\shortauthors{Melatos et al.}

\begin{document}
\title{Measuring the vortex-nucleus pinning force from pulsar glitch rates}

\author{A. Melatos\altaffilmark{1,2} 
 and M. Millhouse \altaffilmark{1,2,3}}

\email{amelatos@unimelb.edu.au}

\altaffiltext{1}{School of Physics, University of Melbourne,
 Parkville, VIC 3010, Australia}

\altaffiltext{2}{Australian Research Council Centre of Excellence
 for Gravitational Wave Discovery (OzGrav)}

\altaffiltext{3}{Center for Relativistic Astrophysics,
 Georgia Institute of Technology, Atlanta, GA 30332, USA}

\begin{abstract}
\noindent 
Superfluid vortex avalanches are one plausible cause of pulsar glitch activity.
If they occur according to a state-dependent Poisson process,
the measured long-term glitch rate is determined by the 
spin-down rate of the stellar crust, $\dot{\Omega}_{\rm c}$,
and two phenomenological parameters quantifying the vortex-nucleus pinning force:
a crust-superfluid angular velocity lag threshold, $X_{\rm cr}$, 
and a reference unpinnng rate, $\lambda_0$.
A Bayesian analysis of 541 glitches in 177 pulsars,
with $N_{\rm g} \geq 1$ events per pulsar,
yields $X_{\rm cr} = 0.15^{+0.09}_{-0.04} \, {\rm rad \, s^{-1}}$,
$\lambda_{\rm ref} = 7.6^{+3.7}_{-2.6} \times 10^{-8} \, {\rm s^{-1}}$,
and $a = -0.27^{+0.04}_{-0.03}$
assuming the phenomenological rate law
$\lambda_0 = \lambda_{\rm ref} [\tau/(1 \, {\rm yr})]^a$,
where $\tau$ denotes the characteristic spin-down age.
The results are broadly similar,
whether one includes or excludes quasiperiodic glitch activity,
giant glitches, or pulsars with $N_{\rm g}=0$,
up to uncertainties about the completeness of the sample and the total observation time
per pulsar.
The $X_{\rm cr}$ and $\lambda_0$ estimates are consistent with first-principles calculations
based on nuclear theory, e.g.\ in the semiclassical local density approximation. 
\end{abstract}


\section{Introduction 
 \label{sec:pin1}}
Neutron stars are natural nuclear laboratories.
Measurements of their masses and radii,
based on X-ray and gravitational-wave data
\citep{mil19,cha20,raa20},
constrain the composition, nucleon-nucleon couplings,
and equation of state of bulk nuclear matter
\citep{lat07,gan15}.
Different stellar strata contain different thermodynamic phases,
ranging from a superfluid in the core and inner crust
\citep{yak99}
to a Coulomb crystal in the outer crust
\citep{cha08}
and amorphous tube-and-sheet `pasta' in between
\citep{cap17}.
At the neutron drip interface,
where the superfluid and crystalline phases coexist,
it is favorable energetically for quantized vortices in the rotating superfluid
to pin at nuclear lattice sites or their interstices
\citep{alp77,jon91}.
Vortex pinning is ubiquitous in terrestrial condensed matter systems,
such as superfluid helium, dilute gas Bose-Einstein condensates, and type II superconductors
\citep{tsa80,fie95,tun06},
but the phenomenon
cannot be studied at nuclear densities in terrestrial laboratories
\citep{gra17}.

Rotational glitches in rotation-powered pulsars offer a unique opportunity
to perform a fundamental nuclear experiment:
a measurement of the vortex-nucleus pinning force $F_{\rm p}$ near the neutron drip density.
Pioneering studies have predicted $F_{\rm p}$ theoretically,
e.g.\ within the semiclassical local density approximation
\citep{alp84a,eps88,don06,avo08,sev16,lin22},
but the calculations are acutely sensitive to quantum shell effects,
screening, and the mesoscopic interaction across multiple pinning sites,
all of which are uncertain without controlled laboratory experiments
\citep{lin91,cao06,gan15,sev16,lin22}.
Indeed even the sign of $F_{\rm p}$ is unknown
\citep{wla16}.
Here it is shown that $F_{\rm p}$ can be inferred from glitch event statistics, 
assuming that glitches are triggered by collective unpinning of superfluid vortices,
i.e.\ vortex avalanches
\citep{and75,war11,has15}.
The key insight is that $F_{\rm p}$ can be related theoretically to
the long-term glitch and spin-down rates,
which are measurable in hundreds of objects at present,
if glitch activity obeys a {\em state-dependent Poisson process}
\citep{ful17}.
The state-dependent Poisson process is a phenomenological, microphysics-agnostic meta-model,
which describes the statistics of threshold-triggered, stick-slip, relaxation events in a 
slowly and continuously stressed system.
It is consistent with the canonical picture of pulsar glitch activity 
as a spin-down-driven phenomenon
and makes falsifiable predictions about glitch observables,
e.g.\ auto- and cross-correlations of waiting times and sizes
\citep{mel18,car19a,car19b,car21}.

The paper is structured as follows.
In \S\ref{sec:pin2}, 
we specify how a state-dependent Poisson process maps onto the microphysics of vortex avalanches
and write down an implied likelihood
for the observed number of glitches per pulsar as a function of the observation time.
In \S\ref{sec:pin3}, 
we perform a Bayesian analysis of 541 glitches detected in 177 pulsars
to infer posterior distributions 
of the phenomenological parameters in the state-dependent Poisson meta-model,
verify the plausibility of the meta-model via informal checks,
and note some important selection effects.
In \S\ref{sec:pin4},
the posterior distributions of the meta-model parameters
are interpreted in terms of microphysical parameters specific to the vortex-nucleus pinning force.

We emphasize at the outset that superfluid vortex avalanches are one of several plausible
mechanisms of pulsar glitch activity,
and that different mechanisms may operate in different pulsars 
or even in the same pulsar simultaneously.
Starquakes \citep{fra00,mid06,chu10b,gil20,ker22}
and hydrodynamic instabilities \citep{and03,gla09} are among the alternatives.
The bimodal distribution of glitch sizes is one possible indicator of multiple mechanisms,
with smaller events associated arguably with starquakes,
and larger events associated with superfluid vortex avalanches
\citep{ash17,fue17,cel20,ant22,zho22,aru23}.
Extending the analysis in this paper to other microphysics may prove fruitful in the future, 
if the approach withstands the tests applied by enlarged data sets
\citep{kra10,cal16}.
For example, the state-dependent Poisson process maps naturally onto the microphysics
of starquakes, with revised interpretations of the meta-model parameters of course.

\section{Glitch rate
 \label{sec:pin2}}
Pulsar glitch activity is often conceptualized as a stress-relax process
\citep{has15,ant22,zho22}.
Stress builds up inside the star, as the crust spins down,
and relaxes sporadically via impulsive, stick-slip events,
when the stress approaches a threshold.
In the starquake and vortex avalanche pictures, for example, 
the stress corresponds physically to crustal elastic deformation
and crust-superfluid differential rotation respectively.
In \S\ref{sec:pin2a}, 
we map the stress-relax dynamics of vortex avalanches onto a state-dependent Poisson process
and express the long-term glitch rate in terms of two phenomenological vortex pinning parameters.
In \S\ref{sec:pin2b}, 
we propose a likelihood function for the observed number of glitches per pulsar,
as well as priors for the phenomenological parameters motivated by nuclear theory
\citep{lin93,sev16}.

\subsection{Vortex avalanches as a state-dependent Poisson process
 \label{sec:pin2a}}
\begin{figure}[ht]
\begin{center}
\includegraphics[width=11cm,angle=0]{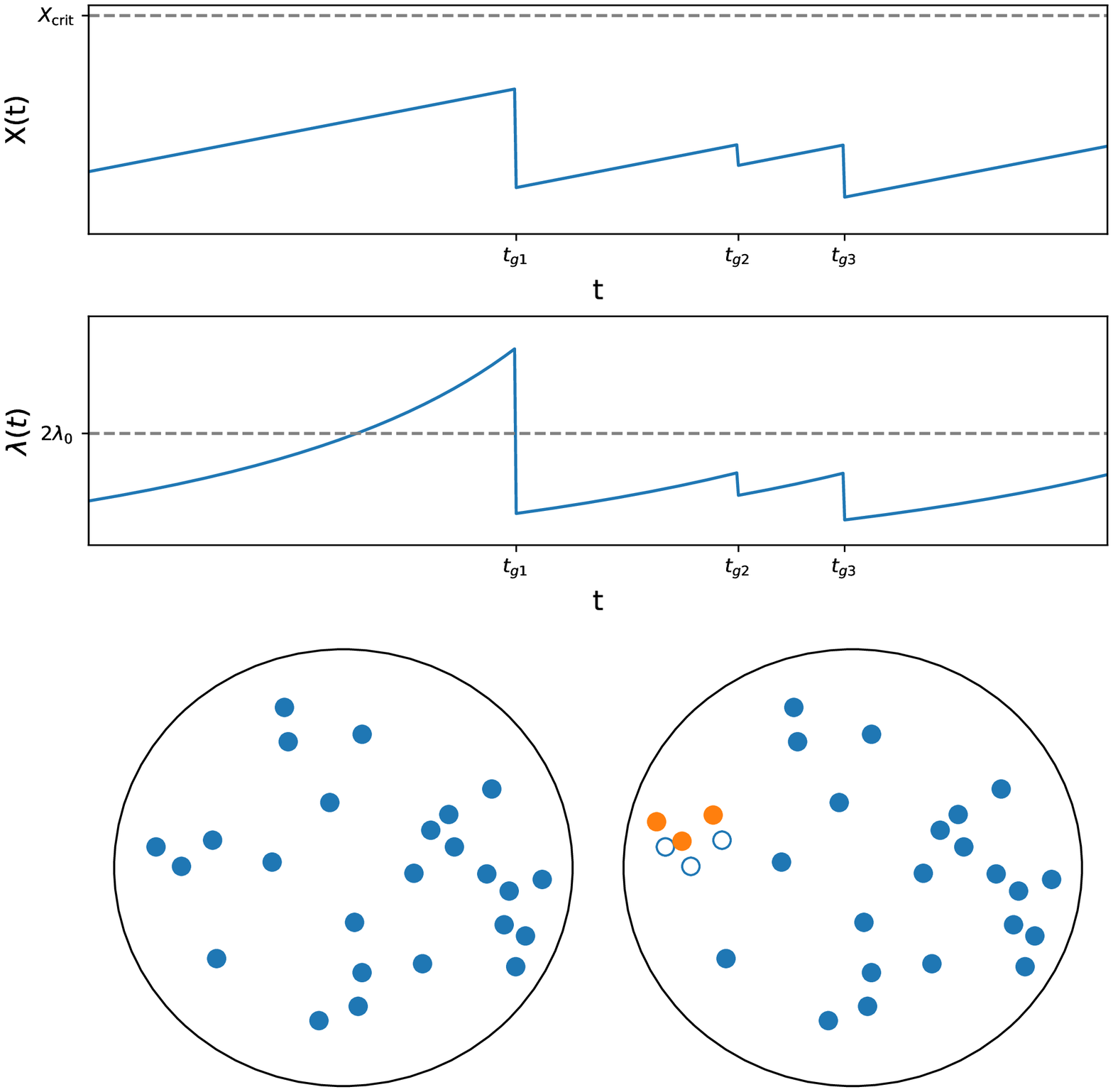}
\end{center}
\caption{
Glitches triggered by superfluid vortex avalanches.
({\em Top panel.})
Angular velocity lag $X(t)$ between the crust and the superfluid versus time $t$. 
The lag increases linearly between the three glitches at $t=t_{\rm g1}$, $t_{\rm g2}$, $t_{\rm g3}$, 
because the superfluid is pinned,
and the crust brakes electromagnetically.
The lag steps down by a random amount at every glitch,
when a vortex avalanche transfers angular momentum from the superfluid to the crust.
({\em Middle panel.}) 
Instantaneous unpinning rate $\lambda(t)=\lambda_0 [ 1 - X(t)/X_{\rm cr} ]^{-1}$ versus time $t$.
The unpinning rate increases with $X(t)$ and diverges for $X(t) \rightarrow X_{\rm cr}$.
({\em Bottom panels.})
Exaggerated schematic of the pinned vortex configuration between glitches, 
drawn in equatorial cross-section.
Pinning traps the vortices metastably in irregular, non-minimum-energy configurations
distinct from a regular Abrikosov array.
The left and right panels display configurations before and after a glitch respectively;
filled circles indicate pinned vortices,
while open circles (in the right panel) indicate positions vacated by three vortices
participating in an avalanche (which pin subsequently at the orange locations).
The departures from a regular Abrikosov array are exaggerated for visual effect.
A typical neutron star contains 
$\approx 3\times 10^{15}(\Omega_{\rm s}/{\rm 1\,rad\,s^{-1}})$
vortices.
}
\label{fig:pin1}
\end{figure}

In the vortex avalanche picture,
the star's rigid crust and superfluid core rotate differentially,
with angular velocities $\Omega_{\rm c}$ and $\Omega_{\rm s} > \Omega_{\rm c}$ respectively.
Glitches occur sporadically, 
when vortices unpin {\em en masse} and move outwards,
transferring angular momentum from the superfluid to the crust.
Figure \ref{fig:pin1} illustrates schematically,
how the lag $X(t) = \Omega_{\rm s}(t) - \Omega_{\rm c}(t)$ evolves
from one glitch to the next.
In between any two glitches,
$X(t)$ increases linearly with time as
$X(t)=X(t_{\rm g1})+ | \dot{\Omega}_{\rm c} | t$
for
$t_{\rm g1} \leq t \leq t_{\rm g2}$
(top panel, Figure \ref{fig:pin1}),
because the pinned superfluid has $\Omega_{\rm s}(t)=\Omega_{\rm s}(t_{\rm g1}) = {\rm constant}$,
and the crust spins down electromagnetically.
At any instant, there is a lag-dependent probability,
that a vortex unpins and triggers a multi-vortex avalanche,
which is modeled phenomenologically by the instantaneous unpinning rate,
$\lambda(t) = \lambda_0 [ 1 - X(t)/X_{\rm cr} ]^{-1}$
\citep{ful17}.
Here $X_{\rm cr}$ is a lag threshold, where unpinning is certain to occur,
and $2\lambda_0$ equals the reference rate at $X(t)=X_{\rm cr}/2$.
Vortex-nucleus pinning controls $\lambda(t)$
through $X_{\rm cr}$ and $\lambda_0$.
The middle panel of Figure \ref{fig:pin1} illustrates how $\lambda(t)$ evolves.
The waiting time $t_{\rm g2} - t_{\rm g1}$
is a random variable obeying an inhomogeneous, state-dependent Poisson process,
whose rate $\lambda(t)$ depends on $X(t)$.
It satisfies
$t_{\rm g2} - t_{\rm g1} \leq [ X_{\rm cr} - X(t_{\rm g1}) ] / | \dot{\Omega}_{\rm c} | $;
every glitch occurs before the lag reaches $X_{\rm cr}$.

When a glitch occurs, say at $t_{\rm g2}$,
vortices unpin in an avalanche,
$\Omega_{\rm s}$ steps down,
$\Omega_{\rm c}$ steps up proportionally to conserve angular momentum,
and $X$ steps down by an amount $\Delta X$,
which depends on the change in $\Omega_{\rm s}$
and the crust and superfluid moments of inertia
\citep{mon20}.
The vortices displaced during an avalanche are illustrated schematically 
in exaggerated form in the bottom panels of Figure \ref{fig:pin1}.
The step $\Delta X > 0$ is a random variable,
whose probability density function (PDF)
$\eta[\Delta X | X(t_{\rm g2}) ]$ is determined by the avalanche microphysics,
e.g.\ knock-on processes observed in quantum mechanical simulations
\citep{war11,war12b},
and is conditional on the lag $X(t_{\rm g2})$ just before the glitch.
Avalanche processes like the one depicted in Figure \ref{fig:pin1},
where stress (here $\Delta X$) accumulates under the action of a slow, global driver
(here electromagnetic braking) and relaxes locally through impulsive, stick-slip events
(here vortex avalanches), are common in nature.
They play a central role in earthquakes
\citep{bur67},
magnetic flux penetration in type II superconductors
\citep{fie95},
eruptive solar phenomena
\citep{whe08,asc18},
and self-organized critical systems in astrophysics more generally
\citep{jen98,asc18}.

A master equation analysis of the state-dependent Poisson process in Figure \ref{fig:pin1}
predicts a relation between two pulsar observables:
the average, long-term glitch rate,
$\gamma = \langle \lambda(t) \rangle$, and the spin-down rate,
$\dot{\Omega}_{\rm c}$. The result is
\citep{ful17}
\begin{equation}
 \gamma = \lambda_0 + \frac{A | \dot{\Omega}_{\rm c} | }{X_{\rm cr}}~,
\label{eq:pin2}
\end{equation}
where $A \sim 1$ is a dimensionless coefficient which describes the shape of
$\eta[\Delta X | X(t_{\rm g2}) ]$,
e.g.\ one has $\eta\propto (\Delta X)^{A-2}$, if $\eta$ is a power law
\citep{ful17}.
We absorb $A$ into $X_{\rm cr}$ henceforth, as the two quantities cannot be disentangled 
observationally; that is, we write $X_{\rm cr}$ instead of $X_{\rm cr}/A$.
Equation (\ref{eq:pin2}) is deceptively simple;
it is the result of a self-consistent, first-principles calculation
\citep{ful17},
which is too lengthy to summarize here.

The theory leading to (\ref{eq:pin2}) reproduces
the observed event statistics of pulsar glitches
in objects with enough data,
including size and waiting time PDFs, auto-, and cross-correlations
\citep{mel18,car19a,car19b}.
In the fast braking limit, the theory also predicts
$\gamma \propto | \dot{\Omega}_{\rm c} |$
through (\ref{eq:pin2}).
The prediction is consistent with the scaling
$\gamma = (4.2 \pm 0.5) \times 10^2 [ | \dot{\Omega}_{\rm c} | / (2\pi \, {\rm Hz\,s^{-1}} ) ]
 \, {\rm s^{-1}}$
for $-13.5 \leq \log[ | \dot{\Omega}_{\rm c} | / (2\pi \, {\rm Hz\,s^{-1}}) ] \leq -10.5$
discovered empirically by other authors
\citep{mck90,lyn00,esp11,fue17,ant18,fer18,ho20};
see Figure 6 in \citet{fue17} for example.
Simulations confirm that
(\ref{eq:pin2}) is insensitive to the exact mathematical forms
of $\lambda(t)$ and $\eta[\Delta X | X(t_{\rm g2}) ]$
\citep{ful17}.

\subsection{Likelihood and priors
 \label{sec:pin2b}}
Hierarchical Bayesian methods 
\citep{gel13}
can be used to infer the nuclear parameters $\lambda_0$ and $X_{\rm cr}$
across the known pulsar population
given per-pulsar measurements of the number of glitches $N_{\rm g}$, 
observation time $T_{\rm obs}$, and $\dot{\Omega}_{\rm c}$.

We define a per-pulsar likelihood function
\begin{equation}
 p[N_{\rm g}^{(k)} | \lambda_0^{(k)}, X_{\rm cr}^{(k)}, \dot{\Omega}_{\rm c}^{(k)}, T_{\rm obs}^{(k)}] 
 = 
 [N_{\rm g}^{(k)} !]^{-1} [\gamma T_{\rm obs}^{(k)} ]^{{N_{\rm g}}^{(k)}}
 \exp[-\gamma T_{\rm obs}^{(k)} ]~,
\label{eq:pin1a}
\end{equation}
as appropriate for a state-dependent Poisson process with average rate
$\gamma[\lambda_0^{(k)},X_{\rm cr}^{(k)}, \dot{\Omega}_{\rm c}^{(k)}]$
\citep{mel19}.
Here $1\leq k \leq K$ indexes the $k$-th pulsar in the sample.
The per-pulsar rate is drawn from a Gaussian PDF, 
\begin{equation}
 \gamma[\lambda_0^{(k)},X_{\rm cr}^{(k)}, \dot{\Omega}_{\rm c}^{(k)}]
 \sim
 {\cal N} [ \lambda_0^{(k)} + | \dot{\Omega}_{\rm c}^{(k)} | / X_{\rm cr}^{(k)}, 
  \sigma ]~, 
\label{eq:pin1b}
\end{equation}
whose mean is given by (\ref{eq:pin2}) with $A$ absorbed into $X_{\rm cr}$,
and whose standard deviation $\sigma$ parametrizes pulsar-to-pulsar variations
in the vortex unpinning parameters caused by stellar mass and composition variations, for example,
which are not captured by the idealized meta-model underpinning (\ref{eq:pin2}).
The likelihood (\ref{eq:pin1a}) is a good approximation for many physically plausible choices of 
$\lambda(t)$ and $\eta[\Delta X | X(t_{\rm g2}) ]$,
including those that generate quasiperiodic glitches in the regime $N_{\rm g} \gg 1$,
where (\ref{eq:pin1a}) approaches a Gaussian
\citep{car19b}.
In this paper, we follow the literature and classify three objects
as glitching quasiperiodically:
PSR J0537$-$6910, PSR J0835$-$4510, and PSR J1341$-$6220,
with $20 \leq N_{\rm g} \leq 45$
\citep{fue17,how18,fue19}.

The nuclear parameters $\lambda_0$ and $X_{\rm cr}$
vary across the pulsar population as functions of hard-to-measure stellar properties
such as the temperature, mass, composition, and superfluid phase
\citep{lin93,sev16}.
One cannot predict the PDFs of $\lambda_0$ and $X_{\rm cr}$
from first principles at the time of writing;
there is too much uncertainty in the nuclear physics
(in the absence of controlled laboratory experiments)
and the astrophysics
(e.g.\ stellar composition).
Therefore much of the pulsar-to-pulsar variation is bundled unavoidably
into the phenomenological dispersion $\sigma$ in (\ref{eq:pin1b}).
However, one can break out the temperature dependence,
under the idealized approximation that the temperature decreases monotonically
as a function of the characteristic spin-down age, 
$\tau = \Omega_{\rm c} / ( 2 | \dot{\Omega}_{\rm c} | )$,
which is observable.
Intuitively, the vortex unpinning rate increases, when thermal fluctuations are greater
\citep{alp84a,lin93},
so $\lambda_0$ increases, as $\tau$ decreases.
We write 
\begin{equation}
 \lambda_0^{(k)} = \lambda_{\rm ref} [ \tau^{(k)} / \tau_{\rm ref} ]^a
\label{eq:pin1c}
\end{equation}
phenomenologically to approximate the latter effect,
with $\tau_{\rm ref} = 1\, {\rm yr}$ without loss of generality.
The parameters $\lambda_{\rm ref}$ and $a$ in (\ref{eq:pin1c})
are to be estimated by the Bayesian analysis in \S\ref{sec:pin3} 
and are assumed to take unique values across the pulsar population.
The existence of an age dependence exemplified by (\ref{eq:pin1c})
is consistent empirically with the Bayesian analysis in \S\ref{sec:pin3},
which rules out $a=0$ with more than 99\% confidence;
see \S\ref{sec:pin3a} specifically as well as \citet{mil22}.
One can argue equally for an age dependence in $X_{\rm cr}$ on analogous intuitive grounds.
However, we assume for simplicity that $X_{\rm cr}^{(k)} = X_{\rm cr}$
takes a unique value across the pulsar population for the purposes of this paper,
while we await more data.
The above approximations involving $\lambda_0$ and $X_{\rm cr}$
are related to the microphysics of the vortex-nucleus pinning force in \S\ref{sec:pin4}.

The challenges in measuring stellar temperature, mass, composition,
and superfluid phase make it difficult to assign informative priors
to $\lambda_{\rm ref}$, $a$, $X_{\rm cr}$, and $\sigma$.
As $\lambda_0$ and $X_{\rm cr}$ involve exponential Boltzmann factors
(e.g.\ through the superfluid energy gap; see \S\ref{sec:pin4}),
it is plausible that they span several decades.
We therefore assume that the priors on $\lambda_{\rm ref}$, $a$, and $X_{\rm cr}$
are broad, with 
$p[\log_{10} (\lambda_{\rm ref} / 1\,{\rm s^{-1}}) ]
 \sim {\cal U}(-30,1)$ (log-uniform),
$p(a) \sim {\cal U}(-3,3)$,
and $p[ (X_{\rm cr} / 1 \, {\rm rad \, s^{-1}} )^{-1} ] \sim {\cal U}(0,\infty)$,
consistent with nuclear theory; see \S\ref{sec:pin4}
\citep{sev16}.
Here, ${\cal U}(x_1,x_2)$ denotes a uniform PDF on the interval $(x_1,x_2)$.
We also assume zero cross-correlation, i.e.\ 
$p(\lambda_{\rm ref},a,X_{\rm cr},\sigma)
 = p(\lambda_{\rm ref})p(a)p(X_{\rm cr}) p(\sigma)$,
in the absence of prior knowledge to the contrary.
A plausible, standard choice for the prior on $\sigma$ is the Jeffreys prior
$p(\sigma) \propto \sigma^{-1}$ 
for the standard deviation of a Gaussian PDF
(equivalent to a log-uniform prior).
In practice, while we await more data,
we adopt a conservative posture in \S\ref{sec:pin3}
and analyze (\ref{eq:pin1a}) and (\ref{eq:pin1b}) in the regime $\sigma \ll \gamma$,
where (\ref{eq:pin1b}) approaches a delta function,
and $\sigma \rightarrow 0$ drops out.
By replacing the $2K$ parameters $\lambda_0^{(k)}$ and $X_{\rm cr}^{(k)}$
with the three parameters $\lambda_{\rm ref}$, $a$, and $X_{\rm cr}$,
we convert the full analysis of (\ref{eq:pin1a}) and (\ref{eq:pin1b})
into a simpler analysis,
which befits the data volume available
\citep{ras06}.
In addition, for the sake of completeness,
we present in Appendix \ref{sec:pinappa} an extended analysis,
where $\sigma$ is estimated conditional on the Jeffreys prior,
without assuming $\sigma \ll \gamma$.
The extended analysis hints at what may be learned physically with more data in the future, 
e.g.\ measuring age-independent pulsar-to-pulsar variations in $\lambda_0$ and $X_{\rm cr}$
through $\sigma$.
We emphasize that the results in Appendix \ref{sec:pinappa} are preliminary
and must be revisited, when more data become available
\citep{kra10,cal16}.

\section{State-dependent Poisson process parameters
 \label{sec:pin3}}
The likelihood and priors in \S\ref{sec:pin2b} can be combined with the observed
event statistics of glitches across the pulsar population to estimate 
$\lambda_{\rm ref}$, $a$, and $X_{\rm cr}$.
In \S\ref{sec:pin3a}--\S\ref{sec:pin3d}
we analyze the subsample of $K=177$ pulsars with $N_{\rm g} \geq 1$,
registered in the Jodrell Bank Observatory and
Australia Telescope National Facility (ATNF) glitch databases
\citep{esp11,yu13,bas21},
\footnote{
Up-to-date glitch catalogs can be found on-line at
{\tt http://www.jb.man.ac.uk/pulsar/glitches.html}
and
{\tt https://www.atnf.csiro.au/research/pulsar/psrcat}.
}
as well as the full sample of $410$ objects with $N_{\rm g}\geq 0$.
To facilitate comparison,
we analyze exactly the same data set as \citet{mil22}.
Accurate measurements of $\dot{\Omega}_{\rm c}$ are drawn from the
ATNF Pulsar Catalogue
\citep{man05}.
$T_{\rm obs}$ is taken to be the time from discovery to MJD 58849,
i.e.\ the beginning of calendar year 2020, to match \citet{mil22}.
An important caveat is that glitch catalogs may be incomplete for small glitches 
due to lengthy monitoring gaps,
thereby underestimating $N_{\rm g}$ and/or overestimating $T_{\rm obs}$
\citep{jan06,yu17,fue17,mel20,bas21,mil22}.
Completeness is discussed in \S\ref{sec:pin3e}.
Parameter estimation is carried out via a Markov chain Monte Carlo analysis
using the programming language Stan
\footnote{
Stan Development Team (2022),
Stan Modeling Language Users Guide and Reference Manual, version 2.30, 
{\tt https://mc-stan.org}.
}
\citep{car17}
and the {\tt cmdstanpy} interface.
\footnote{
Stan Development Team (2021),
CmdStanPy, version 1.0.4,
{\tt https://pypi.org/project/cmdstanpy}.
}

\subsection{Posterior PDF
 \label{sec:pin3a}}
\begin{figure}[ht]
\begin{center}
\includegraphics[width=16cm,angle=0]{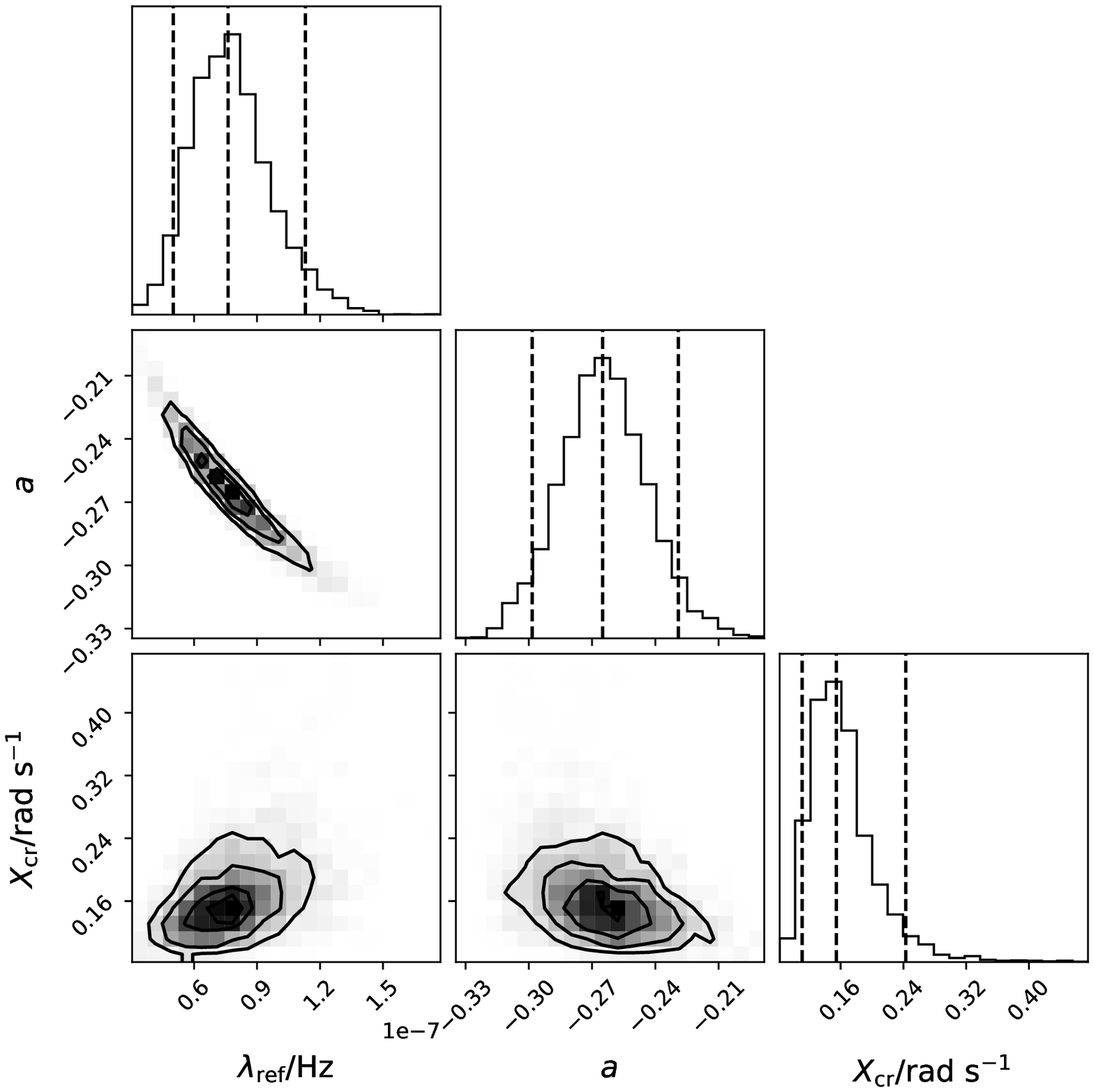}
\end{center}
\caption{
Posterior PDF of the phenomenological parameters
$\lambda_{\rm ref}$ (units: ${\rm s^{-1}}$), 
$a$ (dimensionless), 
and $X_{\rm cr}$ (units: ${\rm rad \, s^{-1}}$)
in the state-dependent Poisson model
for pulsars with $N_{\rm g} \geq 1$,
assuming $\sigma \ll \gamma$.
The layout matches a traditional corner plot, 
with one- and two-dimensional cross-sections of the PDF 
presented as contour plots (with shading) and histograms (solid curves) respectively.
The dashed, vertical lines mark the median and 90\% confidence interval.
}
\label{fig:pin2}
\end{figure}

Figure \ref{fig:pin2} displays the posterior PDF for
$\lambda_{\rm ref}$, $a$, and $X_{\rm cr}$
for objects with $N_{\rm g} \geq 1$ as a traditional corner plot
drawn with the {\tt corner.py} Python module
\citep{for16}.
The PDF is unimodal,
with $\lambda_{\rm ref} = 7.6^{+3.7}_{-2.6} \times 10^{-8} \, {\rm s^{-1}}$,
$a= -0.27^{+0.04}_{-0.03}$,
and
$X_{\rm cr} = 0.15^{+0.09}_{-0.04} \, {\rm rad\,s^{-1}}$
(median and $90\%$ confidence interval).
Widening the priors does not change the PDF.
Upon inspection, there are no strong covariances in Figure \ref{fig:pin2},
except for the familiar trade-off between the amplitude and exponent of a power law, 
which stems from (\ref{eq:pin1c});
as $a$ grows more negative, $\lambda_{\rm ref}$ increases to compensate.

The results in Figure \ref{fig:pin2} assume that $\lambda_0$
varies from pulsar to pulsar through its dependence on $\tau$.
We examine this important assumption independently by temporarily assuming the opposite and
estimating $\lambda_{\rm ref}$ and $X_{\rm cr}$
under the constraint $a=0$.
This is not merely a statistical check;
it is a plausible theoretical hypothesis,
if $\lambda_0$ is determined mainly by $\tau$-independent stellar properties 
which do not vary greatly from pulsar to pulsar, 
such as mass and composition; cf. \S\ref{sec:pin4}.
As in Figure \ref{fig:pin2},
the sampler returns a unimodal PDF,
but the rate PDF generated from the medians of
$\lambda_{\rm ref}$ and $X_{\rm cr}$
overestimates the number of infrequently glitching pulsars 
with $\gamma \lesssim 3\times 10^{-9} \, {\rm s^{-1}}$.
The finding is broadly consistent with a formal Bayesian model selection study
by \citet{mil22},
which concludes that an age-dependent rate law of the form $\gamma \propto \tau^{-0.27}$
is preferred over alternatives proportional to other powers of 
$\Omega_{\rm c}$ and $\dot{\Omega}_{\rm c}$
(including $a=0$)
with log Bayes factor $\geq 3.7$;
see Table 2 in \citet{mil22}.

The $X_{\rm cr}$ estimate in Figure \ref{fig:pin2} is greater than the value 
$(1.5 \pm 0.2)\times 10^{-2} \, {\rm rad \, s^{-1}}$
implied by the scaling 
$\gamma = (4.2 \pm 0.5) \times 10^2 [ | \dot{\Omega}_{\rm c} | / (2\pi \, {\rm Hz\,s^{-1}} ) ]
 \, {\rm s^{-1}}$
for $-13.5 \leq \log[ | \dot{\Omega}_{\rm c} | / (2\pi \, {\rm Hz\,s^{-1}}) ] \leq -10.5$
discovered empirically by other authors
\citep{mck90,lyn00,esp11,fue17,ant18,fer18,ho20}.
However, the results are consistent for two reasons.
First, they are derived from different subsamples.
For example, \citet{fue17} analyzed large glitches (with size $\geq 10\,\mu{\rm Hz}$)
from pulsars satisfying
$-13.5 \leq \log[ | \dot{\Omega}_{\rm c} | / (2\pi \, {\rm Hz\,s^{-1}}) ] \leq -10.5$.
The latter authors also binned the events by spin-down rate,
so that any bin contains events from several pulsars in general
but not all the events from those pulsars (due to the size threshold);
see Figure 6 in \citet{fue17}.
In contrast, Figure \ref{fig:pin2} analyzes all events 
from all objects with $N_{\rm g} \geq 1$ without binning,
irrespective of $\dot{\Omega}_{\rm c}$ and glitch size,
in line with the standard approach in Bayesian inference.
Second, \citet{fue17} were not obliged to include the term $\lambda_0$ in (\ref{eq:pin2}),
because they measured the high-$|\dot{\Omega}_{\rm c}|$ rate law empirically,
whereas (\ref{eq:pin2}) follows from theory
\citep{ful17}.
The $\lambda_0$ term contributes to $\gamma$ to some extent even in the 
high-$|\dot{\Omega}_{\rm c}|$ regime and shifts the Bayesian estimate of $X_{\rm cr}$.
It is important to emphasize that both estimates are valid given current knowledge.
If the glitch activity in all pulsars obeys (\ref{eq:pin2}),
then a subsample without size or $\dot{\Omega}_{\rm c}$ restrictions is more informative.
On the other hand, if a subset of pulsars or glitch sizes does not obey (\ref{eq:pin2}),
perhaps because the $\lambda_0$ term is absent from that subset,
then the large-size, large-$|\dot{\Omega}_{\rm c}|$ subsample is more suitable for that subset.
There is no way to distinguish between the two possibilities with the data and theory available at present.

\subsection{Informal model checks
 \label{sec:pin3b}}
Formal Bayesian model selection between the state-dependent Poisson model 
(\ref{eq:pin2})--(\ref{eq:pin1c}) and competing alternatives,
through a Bayes factor,
falls outside the scope of this paper for two reasons.
First, 
it is computationally prohibitive for the hierarchical analysis in \S\ref{sec:pin3a}
and requires more data than are currently available to achieve clear-cut selection outcomes.
Second,
suitable first-principles competitors to (\ref{eq:pin2})--(\ref{eq:pin1c}) 
do not exist at present,
and non-first-principles alternatives cannot be related to the vortex-nucleus pinning force,
which is the central motivation of this paper.
\footnote{
One first-principles exception, the Brownian stress accumulation model,
does not come with a simple analytic rate formula for $\gamma(\lambda_0,X_{\rm cr})$ 
like (\ref{eq:pin2}) at the time of writing
\citep{car20}.
}
Accordingly, we turn to informal checks,
a common approach \citep{gel13},
to assess the plausibility of the model.

\begin{figure}[ht]
\begin{center}
\includegraphics[width=16cm,angle=0]{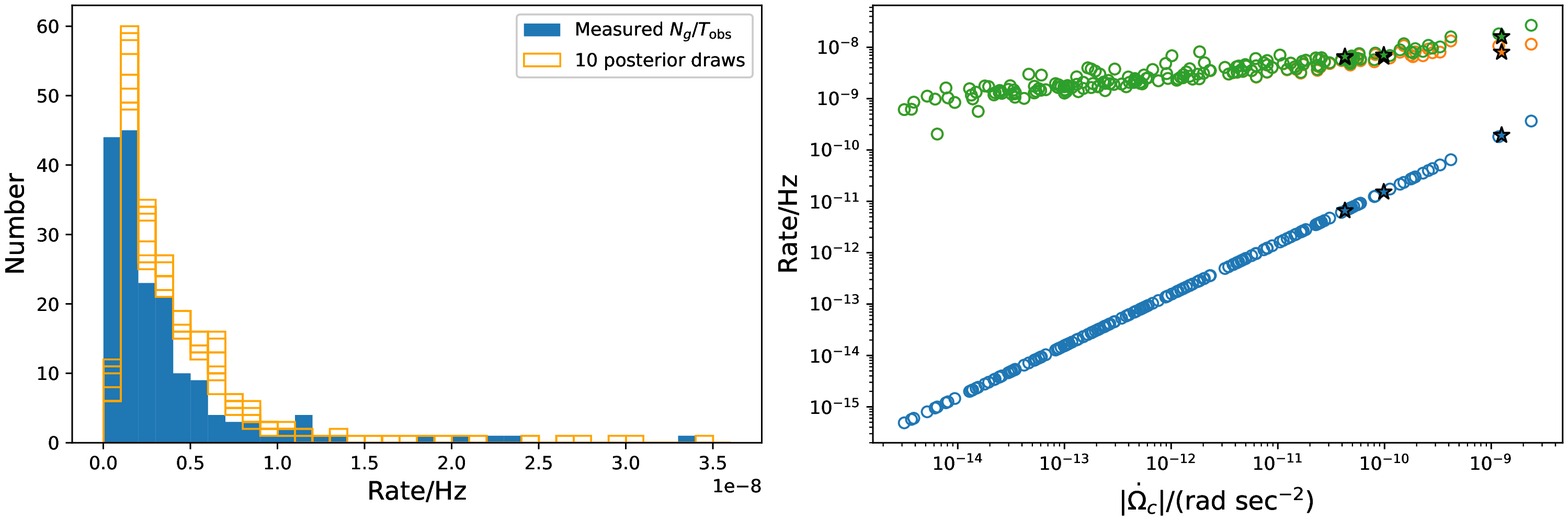}
\end{center}
\caption{
Informal checks of the model (\ref{eq:pin2})--(\ref{eq:pin1c})
with the median estimates for 
$\lambda_{\rm ref}$, $a$, and $X_{\rm cr}$
in Figure \ref{fig:pin2},
assuming $\sigma \ll \gamma$.
({\em Left panel.})
Predicted (orange histograms) and observed (blue histogram) PDFs
of the mean glitch rate $\langle \gamma \rangle = N_{\rm g} / T_{\rm obs}$.
There are $10$ orange histograms,
corresponding to 10 random draws from the posterior PDF in Figure \ref{fig:pin2}.
({\em Right panel.})
Contributions to the per-pulsar rate $\gamma$ (units: ${\rm s^{-1}}$; green circles) 
from the first (orange circles) and second (blue circles) terms in (\ref{eq:pin2})
as functions of the spin-down rate $|\dot{\Omega}_{\rm c}|$
for the 177 pulsars with $N_{\rm g} \geq 1$.
The first term in (\ref{eq:pin2}) dominates except for
$|\dot{\Omega}_{\rm c}| \geq 2.4 \times 10^{-9} \, {\rm s^{-2}}$.
The three objects that glitch quasiperiodically are plotted as stars
(green, orange, and blue per object, as for the circles).
Most green and orange symbols overlap closely and are hard to separate by eye.
}
\label{fig:pin3}
\end{figure}

We start by testing how well 
(\ref{eq:pin2})--(\ref{eq:pin1c}) 
regenerate the data when combined with the parameter estimates in Figure \ref{fig:pin2}.
The left panel of Figure \ref{fig:pin3} displays $10$ orange histograms,
each of which is generated by drawing
$\lambda_{\rm ref}$, $a$, and $X_{\rm cr}$
randomly from the posterior PDF in Figure \ref{fig:pin2}
and evaluating and binning $\gamma$ from (\ref{eq:pin2})--(\ref{eq:pin1c})
for the 177 pulsars with $N_{\rm g} \geq 1$.
The histogram of the observed mean rate $\langle \gamma \rangle = N_{\rm g} / T_{\rm obs}$
for the same objects is overplotted in blue.
Qualitatively speaking, the blue and orange histograms are similar,
implying that (\ref{eq:pin2})--(\ref{eq:pin1c}) represent a plausible model.
However, the blue and orange histograms are not the same in detail.
A Kolmogorov-Smirnov test returns strong evidence,
that the blue and orange histograms are drawn from different distributions, 
with p-values ranging from $1.4\times 10^{-4}$ to $4.2\times 10^{-3}$
for the 10 draws.
In particular,
the model underpredicts low rates;
the 10 draws return between six and 12 objects with $\gamma \leq 1\times 10^{-9} \, {\rm s^{-1}}$
versus $44$ observed.
None of this is surprising;
the state-dependent Poisson meta-model is highly idealized,
the data are likely to be contaminated by multiple selection effects
including incompleteness
\citep{esp11,fue19,low21} (see also \S\ref{sec:pin3e}),
and $\langle \gamma \rangle$ is a convenient but crude summary statistic
which assists with the visual comparison in the left panel of Figure \ref{fig:pin3}
but comes with a significant standard error for $N_{\rm g} \sim 1$
and may be biased systematically (especially for $N_{\rm g} \sim 1$)
by the $T_{\rm obs}$ uncertainties discussed in \S\ref{sec:pin3e}.

We also test whether the data span a wide enough range in $| \dot{\Omega}_{\rm c} |$
to constrain both the first and second terms in (\ref{eq:pin2}),
which involve $\lambda_{\rm ref}$ and $X_{\rm cr}$ respectively.
The right panel of Figure \ref{fig:pin3} plots the per-pulsar rate $\gamma$
generated from the posterior
(green circles)
and the contributions from the first and second terms 
(orange and blue circles respectively)
as functions of $| \dot{\Omega}_{\rm c} | $.
The plot is constructed by setting 
$\lambda_{\rm ref}$, $a$, and $X_{\rm cr}$
to their median values from Figure \ref{fig:pin2}.
The first term dominates the second for most objects.
The second term contributes more than $0.5\gamma$ for only one object,
which has
$| \dot{\Omega}_{\rm c} | = 2.4 \times 10^{-9} \, {\rm s^{-2}}$,
and whose second term contributes $0.55\gamma = 1.5 \times 10^{-8} \, {\rm s^{-1}}$.
This explains why the data are sufficient to constrain (\ref{eq:pin2})--(\ref{eq:pin1c})
but not more complicated models, 
where the sampler fails to converge,
e.g.\ models whose second term depends on age
according to $X_{\rm cr} = X_{\rm ref} [ \tau^{(k)} / \tau_{\rm ref} ]^b$.
It also bears on the comparison in \S\ref{sec:pin3a}
between the values of $X_{\rm cr}$ inferred in Figure \ref{fig:pin2}
and implied by the empirical, high-$| \dot{\Omega}_{\rm c} | $ relation
discovered by other authors, 
e.g.\ \citet{fue17}.
Both estimates are uncertain because they rely on a handful of objects
with high $| \dot{\Omega}_{\rm c} | $.

\subsection{Quasiperiodic glitch activity
 \label{sec:pin3c}}
The pulsars that glitch quasiperiodically contribute 88 out of 541 events
to the $N_{\rm g} \geq 1$ subsample.
There are three of them,
with
$-11.2 \leq \log[ | \dot{\Omega}_{\rm c} | / (2\pi \, {\rm Hz\,s^{-1}}) ] \leq -9.70$.
Upon redoing the analysis in \S\ref{sec:pin3a} with
PSR J0537$-$6910, PSR J0835$-$4510, and PSR J1341$-$6220 excluded,
we obtain
$\lambda_{\rm ref} = 6.1^{+3.0}_{-2.0} \times 10^{-8} \, {\rm s^{-1}}$,
$a= -0.25^{+0.03}_{-0.03}$,
and
$X_{\rm cr} = 0.63^{+3.2}_{-0.36} \, {\rm rad\,s^{-1}}$
(median and $90\%$ confidence interval),
in line with Figure \ref{fig:pin2}.
The greatest shift occurs in $X_{\rm cr}$, which is not surprising.
The three quasiperiodic pulsars have relatively high $|\dot{\Omega}_{\rm c}|$ values
and contribute meaningfully to the second term in (\ref{eq:pin2});
see the starred objects in the right panel of Figure \ref{fig:pin3}.
The result suggests that approximating the likelihood with the Poisson formula
(\ref{eq:pin1a}) is acceptable for quasiperiodic glitch activity,
because (\ref{eq:pin1a}) approaches a Gaussian for $N_{\rm g} \geq 20$,
as foreshadowed in \S\ref{sec:pin2b}.
More data are required to answer more subtle questions,
such as whether the quasiperiodic pulsars obey a rate law other than (\ref{eq:pin2})
or can be distinguished physically from the rest of the subsample in some other way.

Two pulsars that glitch quasiperiodically, 
PSR J0537$-$6910 and PSR J0835$-$4510,
are members of a wider category of objects,
that exhibit giant glitches with average absolute event size
$\langle \Delta\Omega_{\rm c} \rangle/(2\pi) \geq 10^{-5} \, {\rm Hz}$
\citep{esp11,ash17,fue17,eya19,cel20,ant22,zho22,aru23}.
It is unclear whether or not giant glitches
are triggered by the same microphysical mechanism as other glitches
\citep{cel20}.
Either way, when giant glitch activity is excluded from the analysis in \S\ref{sec:pin3a},
it is found that the estimates of $\lambda_{\rm ref}$, $a$, and $X_{\rm cr}$
remain broadly consistent with Figure \ref{fig:pin2}
and the results above on quasiperiodic glitch activity.
Details of the giant glitch analysis are recorded in Appendix \ref{sec:pinappb}.

\subsection{Objects with $N_{\rm g}=0$
 \label{sec:pin3d}}
Pulsars with $N_{\rm g}=0$ convey pertinent information about glitch rates,
and hence the vortex-nucleus pinning force in the vortex avalanche picture,
even though they have never been seen to glitch
\citep{esp11,fue17,mil22}.
Intuitively this is because they exclude $\gamma \gg T_{\rm obs}^{-1}$
on a per-pulsar basis with high confidence.
In principle, they should be included in the analysis in \S\ref{sec:pin3a}.
However, it is important to present the $N_{\rm g}\geq 0$ and $N_{\rm g}\geq 1$ analyses
separately, as in other papers, for the following reason.
Bundling together objects with $N_{\rm g} =0$ and $N_{\rm g} \geq 1$ assumes implicitly
that they are identical physically,
in the sense that they all obey (\ref{eq:pin2})--(\ref{eq:pin1c}),
they are all capable of glitching,
and they all glitch eventually as long as they are observed long enough.
This assumption may be false.
An unknown subset of pulsars may never glitch,
due to some unknown physical cause unrelated to (\ref{eq:pin2})--(\ref{eq:pin1c}),
whereupon it would be misleading to group them with other objects.

When the 233 pulsars with $N_{\rm g}=0$
are added to the analysis in \S\ref{sec:pin3a},
the results do not change qualitatively.
The resulting corner plot is not displayed to avoid repetition;
every panel resembles Figure \ref{fig:pin2} closely in shape and character, 
although the scales shift.
The posterior PDF is unimodal and peaks at
$\lambda_{\rm ref} = 5.5^{+2.3}_{-1.6} \times 10^{-7} \, {\rm s^{-1}}$,
$a= -0.48^{+0.03}_{-0.03}$,
and
$X_{\rm cr} = 0.22^{+0.29}_{-0.08} \, {\rm rad \, s^{-1}}$
(median and $90\%$ confidence interval).
The covariance noted in \S\ref{sec:pin3a} between $\lambda_{\rm ref}$ and $a$
remains a feature.
It is clear that including pulsars with $N_{\rm g}=0$ modifies the inferred
parameters significantly,
consistent with other studies 
\citep{esp11,fue17,mil22}.
For example,
$\lambda_{\rm ref}$ and $X_{\rm cr}$ are multiplied by factors of
$\approx 7$ and $\approx 1.4$ respectively relative to the $N_{\rm g} \geq 1$ subsample.
However, the modifications remain within theoretical bounds, 
as discussed in \S\ref{sec:pin4}.

\subsection{Completeness
 \label{sec:pin3e}}
At least two observational biases contaminate the inference study 
in \S\ref{sec:pin3a}--\S\ref{sec:pin3d}.
First, the glitch catalogs may be incomplete, in the sense that they miss some small glitches.
Second, the definition of $T_{\rm obs}$ in \S\ref{sec:pin3} may overestimate $T_{\rm obs}$
for some objects.
Both biases act to underestimate $\gamma$.
To complicate matters, they impact different objects differently.
Both biases have been investigated thoroughly by previous authors
\citep{esp11,fue19,low21}.
It is challenging to correct for them,
because not all of the relevant observational information is disseminated publicly,
e.g.\ complete telescope logs recording observing epochs and cadences for every pulsar.

A detailed study of observational biases lies outside the scope of this paper.
The reader is referred to the careful analyses by other authors 
cited in the previous paragraph for more information.
Instead, in order to illustrate qualitatively the sort of impact which may occur,
we report briefly the results of a simplified worked example.
We repeat the analysis in \S\ref{sec:pin3a},
based on objects with $N_{\rm g} \geq 1$,
by redefining $T_{\rm obs}$ to be the time elapsed from the first glitch to the present
(instead of the discovery epoch to the present),
because some pulsars are monitored sporadically until they glitch,
and events before the first glitch may escape detection.
As long as the observing cadence is adequate to produce a
phase-connected ephemeris spanning $T_{\rm obs}$,
the above definition corresponds to the minimum $T_{\rm obs}$.
The resulting Bayesian analysis yields
$\lambda_{\rm ref} = 6.5_{-2.0}^{+2.7} \times 10^{-7} \, {\rm s^{-1}}$,
$a = -0.48_{-0.02}^{+0.03}$,
and $X_{\rm cr} = 0.22_{-0.09}^{+0.44} \, {\rm rad \, s^{-1}}$.
The revised median estimate of $X_{\rm cr}$ 
is consistent with the error bars ($90\%$ confidence intervals) in \S\ref{sec:pin3a},
shifting by a factor $\approx 1.5$,
while $\lambda_{\rm ref}$ and $a$ shift by factors of $\approx 9$ and $\approx 1.8$ respectively,
indicating a steeper dependence on $\tau$,
when $T_{\rm obs}$ is redefined,
noting the power-law trade-off between $\lambda_{\rm ref}$ and $a$.
We emphasize that this simplified test is included for illustrative purposes
and cannot supplant the careful analyses performed elsewhere 
\citep{esp11,fue19,low21}.
We encourage future pulsar timing campaigns to release complete
telescope logs publicly,
to facilitate the analysis of observational biases and selection effects.

\section{Vortex-nucleus pinning force
 \label{sec:pin4}}
The posterior PDF estimated in Figure \ref{fig:pin2} can be related to
fundamental nuclear physics as follows.
The phenomenological pinning parameters $\lambda_0$ and $X_{\rm cr}$
(up to the factor $A\sim 1$; see \S\ref{sec:pin2a})
are controlled by three mesoscopic quantities, which describe the vortex-nucleus interaction:
the pinning force per unit length, $f_{\rm p}$,
the attack frequency, $\nu_{\rm a}$,
and the activation energy, $E_{\rm a}$.
In this section,
we convert the Bayesian estimates of $\lambda_0$ and $X_{\rm cr}$ from \S\ref{sec:pin3}
into bounds on $f_{\rm p}$ (in \S\ref{sec:pin4a}) 
and $E_{\rm a}$ (in \S\ref{sec:pin4b}),
assuming theoretical predictions of $\nu_{\rm a}$.
We also consider briefly the implications for vortex creep and post-glitch recoveries
in \S\ref{sec:pin4c}.

\subsection{Force per unit length
 \label{sec:pin4a}}
\begin{figure}[ht]
\begin{center}
\includegraphics[width=16cm,angle=0]{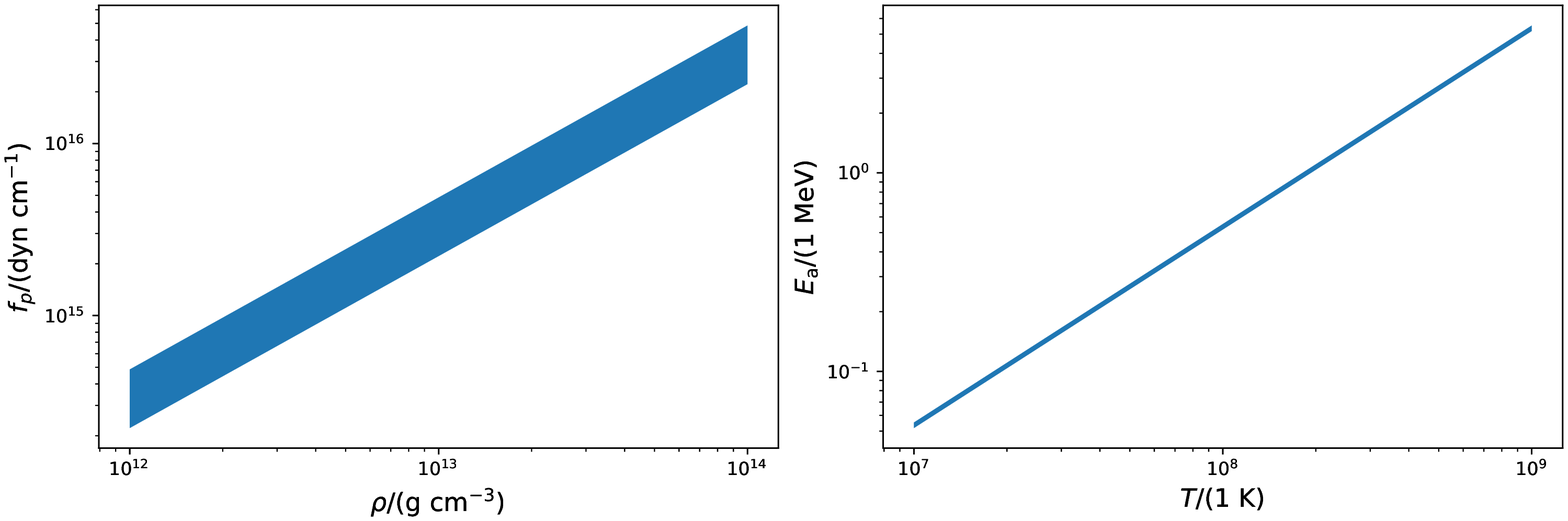}
\end{center}
\caption{
Fundamental nuclear parameters in the vortex-nucleus pinning force,
inferred from the Bayesian estimates of $\lambda_{\rm ref}$, $a$, and $X_{\rm cr}$
in Figure \ref{fig:pin2}.
Allowed values are bounded by the shaded parallelograms.
({\em Left panel.})
Pinning force per unit length, $f_{\rm p}$ (units: ${\rm dyn\, cm^{-1}}$),
in the superfluid density range
$10^{12}\leq \rho  / (1\,{\rm g\,cm^{-3}}) \leq 10^{14}$.
The bottom and top edges of the parallelogram correspond to the 
lower and upper bounds of the
$90\%$ confidence interval of $X_{\rm cr}$ respectively.
({\em Right panel.})
Activation energy, $E_{\rm a}$ (units: ${\rm MeV}$),
in the temperature range
$10^7 \leq T / (1 \, {\rm K}) \leq 10^9$.
The bottom and top edges of the parallelogram correspond to the 
median estimates of $\lambda_{\rm ref}$ and $a$
with $\tau = 10^3 \, {\rm yr}$ and $10^6 \, {\rm yr}$ respectively,
combined with the theoretical bounds
$\nu_{\rm a} = 10^{18} \, {\rm s^{-1}}$
and
$10^{19} \, {\rm s^{-1}}$ respectively.
}
\label{fig:pin4}
\end{figure}

By balancing the hydrodynamic lift force (Magnus force) against the pinning force,
one obtains \citep{alp84a,lin91}
\begin{equation}
 f_{\rm p}
 =
 2 \times 10^{16}
 \left( \frac{\rho}{10^{13} \, {\rm g\,cm^{-3}}} \right)
 \left( \frac{X_{\rm cr}}{1 \, {\rm rad\,s^{-1}}} \right)
 {\rm dyn \, cm^{-1}}
\label{eq:pin3}
\end{equation}
for a star of radius $10\,{\rm km}$,
where $\rho$ is the superfluid mass density.
The median estimate of $X_{\rm cr}$ in Figure \ref{fig:pin2},
combined with (\ref{eq:pin3}), implies
$0.3 \lesssim f_{\rm p}/(10^{15} \, {\rm dyn \, cm^{-1}}) \lesssim 30$
in the density range
$10^{12}\leq \rho  / (1\,{\rm g\,cm^{-3}}) \leq 10^{14}$.
The full range of allowed $f_{\rm p}$ values for
$10^{12}\leq \rho  / (1\,{\rm g\,cm^{-3}}) \leq 10^{14}$,
encompassing the $90\%$ confidence interval of $X_{\rm cr}$,
is contained within the shaded parallelogram in the left panel of Figure \ref{fig:pin4}.
The value of $f_{\rm p}$ implied by the high-$| \dot{\Omega}_{\rm c} |$ rate scaling
in \citet{fue17} lies $\approx 7$ times below the bottom edge of the parallelogram.

The parallelogram in the left panel of Figure \ref{fig:pin4}
is consistent with recent predictions from nuclear theory.
Calculations based on the local density approximation
give
$5\times 10^{15} \lesssim |f_{\rm p}| / (1 \, {\rm dyn \, cm^{-1}}) \lesssim 5\times 10^{17}$
in the density range
$10^{13}\leq \rho  / (1\,{\rm g\,cm^{-3}}) \leq 10^{14}$
irrespective of the sign of $f_{\rm p}$,
with $|f_{\rm p}|$ peaking at $\rho \approx 8\times 10^{13} \, {\rm g\, cm^{-3}}$
\citep{don06,sev16}.
The predicted $f_{\rm p}$ range in the previous sentence 
is calculated for the bare nuclear interaction in a body-centered cubic lattice
and vortices with lengths between $1\times 10^2$ and $5\times 10^3$
Wigner-Seitz radii;
see Figure 10 in \citet{sev16} as well as \citet{lin22}.
The left panel of Figure \ref{fig:pin4}
is also consistent with earlier predictions,
which extend to lower densities
$10^{12}\leq \rho  / (1\,{\rm g\,cm^{-3}}) \leq 10^{14}$
\citep{lin93,don06}.
Physically $f_{\rm p}$ is set by the superfluid energy gap and Fermi momentum
\citep{alp84a,avo08},
modified by local randomness in the nuclear lattice
\citep{jon91,sev16,lin22}.
Theory predicts $f_{\rm p}$ to be $\approx 5$ times lower,
when the polarization of the medium or a random lattice are included;
see Figures 10 and 14 in \citet{sev16} respectively.

A trade-off is made when interpreting $f_{\rm p}$ 
between the intrinsic strength of the nuclear interaction
and the locations, where vortex avalanches occur.
By inferring $f_{\rm p}$ from (\ref{eq:pin3}), 
we assume implicitly that the pinning layer is homogeneous 
and characterized by a unique density $\rho$.
In reality, pinning may occur in multiple layers with different densities,
complicating the interpretation.
For example, the relatively high lag threshold obtained from Figure \ref{fig:pin2},
viz.\ $X_{\rm cr} = 0.15 \, {\rm rad \, s^{-1}}$ (median),
implies either that pinning is relatively strong in the inner crust,
or that vortex avalanches start in the type II superconducting outer core,
where $f_{\rm p}$ is boosted by pinning to magnetic flux tubes,
and vortex creep operates in the nonlinear regime (see \S\ref{sec:pin4c})
\citep{gug16}.
If the outer core does play a role,
then the bare nuclear interaction is modified mesoscopically 
by random tangling between vortices and flux tubes
\citep{dru18},
similarly to how it is modified by local randomness in the nuclear lattice
in the inner crust
\citep{jon91,sev16,lin22}.
On the other hand, if vortex avalanches are confined to the inner crust,
one must interpret $f_{\rm p}$ with crustal entrainment in mind, 
i.e.\ dissipationless coupling between lattice nuclei
and unbound superfluid neutrons in the conduction band
\citep{and12,cha12,cha13}.
Entrainment increases $X_{\rm cr}$
by decreasing the vortex tension and Magnus force;
see equations (29) and (30) in \citet{lin14}.

\subsection{Activation energy
 \label{sec:pin4b}}
A vortex unpins classically by thermal activation or quantum mechanically by tunneling.
The unpinning rate obeys an Arrhenius law,
which relates $\lambda_0 \ll \nu_{\rm a}$ to the activation energy $E_{\rm a}$ via
\citep{alp84a,lin93}
\begin{equation}
 E_{\rm a}
 = 
 0.52
 \left( \frac{T}{10^8 \, {\rm K}} \right)
 \left\{
 1 +
 0.017 \ln\left[
  \left( \frac{\nu_{\rm a}}{10^{18} \, {\rm s^{-1}}} \right)
  \left( \frac{\lambda_0}{10^{-8} \, {\rm s^{-1}}} \right)^{-1}
 \right] 
 \right\} \, {\rm MeV}~,
\label{eq:pin4}
\end{equation}
where $T$ is the superfluid temperature, and $k_{\rm B}$ is Boltzmann's constant.
The logarithm in (\ref{eq:pin4}) means that a measurement of $\lambda_0$
constrains $E_{\rm a}$ more tightly than $\nu_{\rm a}$.
The median estimates of $\lambda_{\rm ref}$ and $a$ in Figure \ref{fig:pin2},
combined with (\ref{eq:pin4}),
put $E_{\rm a}$ in the range
$0.52 \lesssim E_{\rm a}/ (1\,{\rm MeV}) \lesssim 0.54$
for $10^3 \leq \tau / (1 \, {\rm yr}) \leq 10^6$
and the fiducial settings
$T = 10^8 \, {\rm K}$ and $\nu_{\rm a} = 10^{18} \, {\rm s^{-1}}$.
The full range of allowed $E_{\rm a}$ values
for $10^7 \leq T / (1 \, {\rm K}) \leq 10^9$,
assuming the median estimates of $\lambda_{\rm ref}$ and $a$
and $10^3 \leq \tau / (1 \, {\rm yr}) \leq 10^6$,
is contained within the shaded parallelogram in the right panel of Figure \ref{fig:pin4}.
We assume
$10^{18} \leq \nu_{\rm a} / (1 \, {\rm s^{-1}} )
 \leq 10^{19}$
from theoretical calculations,
which include the entropy factor set by the number of vortex vibration modes
\citep{lin93}.
Recent calculations of the vortex attack speed $V_{\rm a}$,
given approximately by $\nu_{\rm a}$ multiplied by the Wigner-Seitz radius, 
yield
$10^5 \lesssim V_{\rm a} / (1 \, {\rm cm\,s^{-1}}) \lesssim 10^7$
\citep{gug16,ant20}.

The parallelogram in the right panel of Figure \ref{fig:pin4}
is broadly consistent with predictions from nuclear theory.
\citet{lin91} found that $E_{\rm a}$ equals $\approx 2.2$ times
the pinning energy per pinning site;
see equation (5.1) in \citet{lin91},
evaluated for $X =  X_{\rm cr}/2$ (as for $\lambda_0$)
and neglecting vortex tension.
The resulting $E_{\rm a}$ values span the range
$4\times 10^{-3} \lesssim E_{\rm a} / (1 \, {\rm MeV}) \lesssim 1\times 10^{1}$
for $10^{12} \leq \rho / (1\, {\rm g\, cm^{-3}}) \leq 10^{14}$;
see Table 1 in \citet{lin91}
and Tables 1 and 2 in \citet{sev16}.
The variation of $E_{\rm a}$ with stellar age enters mainly through $T$,
because $\lambda_0$ is buried inside the logarithm in (\ref{eq:pin4}).
Thermal activation and quantum tunneling dominate for
$k_{\rm B} T \gtrsim  h \nu_{\rm a}/2$ and
$k_{\rm B} T \lesssim h \nu_{\rm a}/2$ respectively, where $h$ is Planck's constant
\citep{lin93}.

Both $E_{\rm a}$ and $\nu_{\rm a}$ are mesoscopic,
as they depend on the number of pinning bonds broken during unpinning,
e.g.\ single- versus multi-site breakaway 
in the regimes of low and high vortex tension respectively.
Although $E_{\rm a}$ and $f_{\rm p}$ are independent,
because $E_{\rm a} / f_{\rm p}$ depends on the number of broken pinning bonds,
a consistency check exists in the low-tension, single-site-breakaway regime.
The mesoscopic pinning force per unit length can be related to
the pinning energy per pinning site, $U_{\rm p}$,
through 
\begin{equation}
 U_{\rm p} 
 =
 0.13 
 \left( \frac{f_{\rm p}}{2\times 10^{16} \, {\rm dyn \, cm^{-1}}} \right)
 \left( \frac{l}{10^{-11} \, {\rm cm}} \right)
 \left( \frac{r_0}{10^{-12} \, {\rm cm}} \right) \, {\rm MeV}~,
\label{eq:pin5}
\end{equation}
where $0.3 \lesssim l/(10^{-11}\,{\rm cm}) \lesssim 1$ is the site separation,
and $0.2 \lesssim r_0/(10^{-12}\,{\rm cm}) \lesssim 2$ is the site radius.
One expects $E_{\rm a} \approx U_{\rm p}$ in the low-tension,
single-site-breakaway regime
$\tau_{\rm v} r_0 / (| f_{\rm p} | l^{2}) \ll 1$,
where $\tau_{\rm v}$ is the vortex tension force,
and one expects $E_{\rm a} > U_{\rm p}$ in the high-tension,
multi-site-breakaway regime
\citep{lin93,sev16}.
Equations (\ref{eq:pin4}) and (\ref{eq:pin5}) satisfy this consistency check
for temperatures $T \gtrsim  10^{7} \, {\rm K}$,
which are reasonable astrophysically,
assuming other variables are held at their fiducial values
by way of illustration.
Once more temperature data are collected, it will be possible in principle
to infer $\tau_{\rm v}$ from the ratio $E_{\rm a} / U_{\rm p}$
and study related phenomena including entrainment (see \S\ref{sec:pin4a}).

\subsection{Vortex creep
 \label{sec:pin4c}}
The vortex-nucleus pinning force plays an important role during post-glitch recoveries,
when fractions of the impulsive jumps $\Delta\Omega_{\rm c}$ and
$\Delta\dot{\Omega}_{\rm c}$ at $t=t_{\rm g1}$ (see \S\ref{sec:pin2a})
are observed to decay quasiexponentially 
during the inter-glitch interval $t_{\rm g1} \leq t \leq t_{\rm g2}$.
It is believed that post-glitch recoveries are governed by vortex creep
\citep{alp84a,lin14},
a type of quasisteady vortex slippage,
in which vortices unpin due to thermal activation, move a short distance, repin,
and then repeat the cycle in a gradual and spatially uncorrelated manner;
cf.\ vortex avalanches \citep{war11}, which are spatially correlated
(see \S\ref{sec:pin2a} and the bottom panels of Figure \ref{fig:pin1}).
It is therefore interesting to check whether the pinning parameters
inferred in \S\ref{sec:pin3} from glitch counts $N_{\rm g}$
are consistent with pulsar timing observations of post-glitch recoveries
interpreted in terms of vortex creep.

In their simplest form,
the recoveries of $\Delta\Omega_{\rm c}$ and $\Delta\dot{\Omega}_{\rm c}$
involve exponential factors $\propto \exp[-(t-t_{\rm g1})/t_{\rm r}]$,
where $t_{\rm r}$ is the superfluid recoupling time,
i.e.\ the time-scale over which superfluid regions affected by
the vortex avalanche at $t_{\rm g1}$
equilibrate with unaffected regions.
\footnote{
Equilibration does not mean corotation with the crust.
The observed absence of strong size-waiting-time cross-correlations
in all glitching pulsars except PSR J0537$-$6910 indicates that the
stress reservoir never empties;
$X(t)$ does not return to zero during the inter-glitch interval 
(see Figure \ref{fig:pin1})
\citep{mel18}.
}
In most pulsars, vortex creep occurs in the nonlinear regime
$\langle X(t) \rangle \lesssim X_{\rm cr}$;
see \citet{alp84a} and {\S}10 in \citet{lin14}.
In the nonlinear regime, the recoupling time is predicted to take the form
\begin{equation}
 t_{\rm r} 
 =
 c_1
 \left( \frac{E_{\rm a}}{k_{\rm B} T} \right)^{-c_2}
 \frac{X_{\rm cr}}{| \dot{\Omega}_{\rm c} |}~,
\label{eq:pin8}
\end{equation}
where $10^{-2} \lesssim c_1 \lesssim 1$
and 
$0.8 \leq c_2 \leq 1$
are dimensionless constants,
whose exact values depend on the microscopic theory in a complicated way;
see equations (32) in \citet{alp84a},
(61) and (92) in \citet{lin14},
and (10) in \citet{gug22}.
The factor $c_1$ is proportional to the crust's moment of inertia
and scales logarithmically with the drag force dissipation angle.
The exponent $c_2$ determines how $t_{\rm r}$ depends on $U_{\rm p}$.
It implies that $t_{\rm r}$ is independent of $U_{\rm p}$ for $c_2=1$,
noting that one has $X_{\rm cr}\propto U_{\rm p} \propto E_{\rm a}$
\citep{lin14}.

The median estimates of $\lambda_{\rm ref}$, $a$, and $X_{\rm cr}$
in Figure \ref{fig:pin2},
combined with (\ref{eq:pin4}), (\ref{eq:pin8}),
and the fiducial choices $c_1=0.1$ and $c_2=1$,
imply
$t_{\rm r} = 2.5 \times 10^{6} 
 (| \dot{\Omega}_{\rm c} | / 10^{-10} \, {\rm rad \, s^{-2}} )^{-1} \, {\rm s}$,
which is of order weeks for 
$| \dot{\Omega}_{\rm c} | \leq 10^{-10} \, {\rm rad \, s^{-2}}$.
Broadly speaking, this is consistent with observations of post-glitch recoveries
\citep{mcc87,lyn92}.
\citet{gug22} recently fitted the vortex creep model to 16 large glitches
in four gamma-ray pulsars:
PSR J0835$-$4510, PSR J1023$-$5746, PSR J2111$+$4606, and PSR J2229$+$6114.
The multi-event fits return 
$19 \leq \tau_{\rm nl} / (1 \, {\rm day}) \leq 45$
and
$59 \leq \tau_{\rm nl} / (1 \, {\rm day}) \leq 91$
for
PSR J0835$-$4510 and PSR J1023$-$5746 respectively,
with $t_{\rm r}$ written as $\tau_{\rm nl}$ by \citet{gug22}.
The fits are in reasonable accord with the PDF in Figure \ref{fig:pin2},
which predicts $t_{\rm r} \approx 29 \, {\rm days}$
and $t_{\rm r} \approx 15 \, {\rm days}$ respectively (median estimates).
\citet{gug22} also fitted a related parameter denoted by $\tau_{\rm s}$,
viz.\ the $e$-folding time observed in the $\Delta\dot{\Omega}_{\rm c}$ timing profile.
The latter fits return
$\tau_{\rm s} = 37 \, {\rm days}$,
$25 \leq \tau_{\rm s} / ( 1\, {\rm day}) \leq 205$,
$127 \leq \tau_{\rm s} / ( 1\, {\rm day}) \leq 143$,
and
$24 \leq \tau_{\rm s} / ( 1\, {\rm day}) \leq 115$
for the four objects respectively.
The $\tau_{\rm s}$ fits are also in reasonable accord with the PDF in Figure \ref{fig:pin2},
which predicts 
$t_{\rm r} \approx 29$, 15, 80, and $16\,{\rm days}$ respectively (median estimates).
Equation (16) in \citet{gug22} implies a theoretical uncertainty of $\approx 1.7$ decades
in $\tau_{\rm nl}$ and hence $t_{\rm r}$,
arising partly from uncertainty about $\rho$ in the recoupling superfluid regions;
see also \citet{gug20}.

Besides post-glitch recoveries, vortex creep also plays an important role
in how $X(t)$ grows between glitches, and how $X(t)$ behaves near $X_{\rm cr}$.
We discuss the two effects in turn.
In the state-dependent Poisson model developed by \citet{ful17},
and in this paper,
we assume $X(t) = X(t_{\rm g1}) + | \dot{\Omega}_{\rm c} | t$
for $t_{\rm g1} \leq t \leq t_{\rm g2}$,
because the assumption is consistent with measurements of size and waiting time PDFs, 
auto-, and cross-correlations
\citep{mel18,car19a,car19b}.
Vortex creep theory can be generalized to handle a time-dependent external torque;
see {\S}4 in \citet{gug17},
where the creep time-scales (e.g.\ $t_{\rm r}$) are calculated in terms of
a running temporal average of the torque.
Likewise,
the analytic theory of the state-dependent Poisson process can be generalized to handle
a time-dependent external torque;
see Appendix E in \citet{ful17}.
The average long-term glitch rate $\gamma$ still obeys (\ref{eq:pin2}) approximately,
but the effective value of $A\sim 1$ is modified somewhat,
and $\dot{\Omega}_{\rm c}$ should be reinterpreted as a running temporal average,
by analogy with \citet{gug17}.
The benefit of keeping the simple formula (\ref{eq:pin2})
to implement the hierarchical Bayesian procedure in \S\ref{sec:pin2b},
especially as (\ref{eq:pin2}) is consistent with measured event statistics
\citep{mel18,car19a,car19b},
outweighs the benefit of evolving the torque,
given the many other theoretical uncertainties in the problem.

The behavior of $X(t)$ near $X_{\rm cr}$ raises some
interesting theoretical issues, which we flag briefly here.
The state-dependent Poisson model maintains $X(t) < X_{\rm cr}$
by construction, because $\lambda[X(t)]$ diverges for $X(t) \rightarrow X_{\rm cr}$.
Likewise, vortex creep maintains $X(t) \lesssim X_{\rm cr}$,
because the radial creep speed rises sharply (without diverging formally)
for $X(t) \rightarrow X_{\rm c}$;
see equation (17) in \citet{alp84a} for example.
However, in the vortex creep picture,
$X(t)$ can fluctuate above $X_{\rm cr}$ temporarily
in response to random stimuli, e.g.\
crustquakes \citep{alp96},
proximity effects in crowded vortex traps \citep{che88},
or thermal pulses \citep{lin96}.
Accordingly, the stress-relax dynamics alter 
to resemble a Brownian stress accumulation process,
where $X(t)$ executes an upwardly biased random walk between glitches
(drift driven by spin down, diffusion driven by vortex creep),
and $\gamma$ is determined by the mean Brownian hitting time at $X(t) = X_{\rm cr}$
\citep{car20}.
One can calculate $\gamma$ analytically
[viz.\ the reciprocal of equation (13) in \citet{car20}], 
but the formula is more complicated than (\ref{eq:pin2}),
making it hard to apply the hierarchical Bayesian procedure in \S\ref{sec:pin2b}.
Moreover, $\gamma$ is a function of the dimensionless parameter
$|\dot{\Omega}_{\rm c} | X_{\rm cr} / \sigma_{\rm ext}^2$,
where $\sigma_{\rm ext}^2$ is the variance of the white-noise driver perturbing $X(t)$,
so the estimation procedure in \S\ref{sec:pin2b} cannot disentangle $X_{\rm cr}$ and $\sigma_{\rm ext}$.
Finally,
Brownian stress accumulation is inconsistent with the event statistics
of some (but not all) pulsars,
unlike the state-dependent Poisson process
\citep{car20}.
\footnote{
Specifically,
PSR J1740$-$3015 has a forward size-waiting-time cross-correlation consistent
with zero and a power-law size PDF. 
This combination is only possible within the Brownian stress accumulation model, 
if the waiting time PDF is a power law too, but instead it is an exponential
\citep{car20}.
The same logic applies to PSR J0631$+$1036.
These conclusions attract the caveat,
that glitch catalogs may be incomplete with respect to small, closely spaced glitches;
see \S\ref{sec:pin3e}
\citep{esp11,fue19,low21}.
}
Consequently, we analyze the state-dependent Poisson process
and its rate law (\ref{eq:pin2}) in this paper
and defer the study of the interesting theoretical issues above to future work.

\section{Conclusion
 \label{sec:pin5}}
The central idea in this paper is that,
if superfluid vortex avalanches in a neutron star obey a state-dependent Poisson process,
the measured glitch rate $\gamma$ relates directly to the phenomenological pinning parameters
$\lambda_0$ and $X_{\rm cr}$ 
through (\ref{eq:pin2})
and hence the fundamental nuclear parameters $f_{\rm p}$ and $E_{\rm a}$
through (\ref{eq:pin3}) and (\ref{eq:pin4}) respectively.
The high-$|\dot{\Omega}_{\rm c}|$ end of (\ref{eq:pin2}) is consistent with
the $\gamma \propto |\dot{\Omega}_{\rm c}|$ scaling discovered empirically
by other authors
\citep{mck90,lyn00,esp11,fue17,ant18,fer18,ho20}.
A Bayesian analysis of 541 glitches in 177 pulsars,
with $N_{\rm g} \geq 1$ events per pulsar,
yields $X_{\rm cr} = 0.15^{+0.09}_{-0.04} \, {\rm rad \, s^{-1}}$,
$\lambda_{\rm ref} = 7.6^{+3.7}_{-2.6} \times 10^{-8} \, {\rm s^{-1}}$,
and $a = -0.27^{+0.04}_{-0.03}$,
assuming the phenomenological rate law
$\lambda_0 = \lambda_{\rm ref} (\tau/\tau_{\rm ref})^a$
with
$\tau_{\rm ref} = 1\, {\rm yr}$.
The median estimates translate into bounds on $f_{\rm p}$ and $E_{\rm a}$,
as expressed through the shaded parallelograms in Figure \ref{fig:pin4},
with
$0.3 \lesssim f_{\rm p}/(10^{15} \, {\rm dyn \, cm^{-1}}) \lesssim 30$
for
$10^{12}\leq \rho  / (1\,{\rm g\,cm^{-3}}) \leq 10^{14}$
and
$0.05 \lesssim E_{\rm a} / ( 1\, {\rm MeV}) \lesssim 5$
for
$10^7 \leq T / (1 \, {\rm K}) \leq 10^9$.
They also translate into bounds on the nuclear parameters governing
vortex creep, e.g.\ during post-glitch recoveries,
as discussed in \S\ref{sec:pin4c}
\citep{alp84a,lin14,gug17,gug20,gug22}.
The bounds are consistent with predictions from nuclear theory
in the local density approximation for a range of lattice structures
(e.g.\ body-centered cubic, random),
vortex lengths
(e.g.\ from $1\times 10^2$ to $5\times 10^3$ Wigner-Seitz radii),
medium polarizations, 
and vortex tension force strengths
\citep{lin91,lin93,don06,sev16,lin22}.
Measurements of the vortex-nucleus pinning force 
in bulk matter near the neutron drip density,
such as the measurements in this paper,
cannot be made in terrestrial laboratories at present.

The Bayesian analysis returns broadly similar results when applied to
physically justified subsamples of the data.
For example,
if the three pulsars that exhibit quasiperiodic glitch activity
(PSR J0537$-$6910, PSR J0835$-$4510, and PSR J1341$-$6220,
with 88 glitches between them)
are excluded from the original analysis,
the median estimates of $\lambda_{\rm ref}$, $a$, and $X_{\rm cr}$
shift by factors $\approx 0.8$, $\approx 0.9$, and $\approx 4$ respectively.
Similar results are obtained,
when the 41 pulsars that exhibit giant glitches are excluded,
as discussed in Appendix \ref{sec:pinappb}.
Likewise,
if the 233 pulsars with $N_{\rm g}=0$ 
\citep{esp11,fue17,mil22} 
are added to the original analysis,
the median estimates of $\lambda_{\rm ref}$, $a$, and $X_{\rm cr}$
shift by factors $\approx 7$, $\approx 1.8$, and $\approx 1.4$ respectively,
noting that a power-law trade-off exists between $\lambda_{\rm ref}$ and $a$.
It is important to emphasize again,
that objects with $N_{\rm g} =0$ and $N_{\rm g} \geq 1$
may or may not be fundamentally different physically;
data available at present cannot resolve this uncertainty.
An unknown subset of pulsars may never glitch,
due to some physical cause unrelated to (\ref{eq:pin2})--(\ref{eq:pin1c});
alternatively, all pulsars may glitch eventually, if one waits long enough.
Other important sources of uncertainty include
the completeness of glitch catalogs to small glitches
and per-pulsar reports of $T_{\rm obs}$.
We encourage future pulsar timing campaigns to publish
detailed per-pulsar logs of observing times and cadences
in a readily searchable form,
to assist with computing $T_{\rm obs}$
and quantifying statistical biases and selection effects.

Pulsar-to-pulsar variations in internal properties that depend weakly on age,
such as mass and chemical composition,
produce some dispersion in $\gamma$ over and above the age dependence
$\lambda_0 \propto \tau^a$ in (\ref{eq:pin1c}).
A preliminary estimate 
$\sigma = 1.1^{+0.5}_{-0.4} \times 10^{-9} \, {\rm s^{-1}}$ 
of the age-independent dispersion is computed in Appendix \ref{sec:pinappa}.
It is consistent with $\sigma \lesssim \gamma$;
that is, age-independent variations are subordinate across the pulsar population.
We emphasize that this result is preliminary,
partly because more data are needed,
and  partly because it is hard to interpret $\sigma$ physically in (\ref{eq:pin1b}),
in light of the many theoretical uncertainties.
However, $\sigma$ is an important quantity.
Expanded glitch catalogs from the next generation of radio timing campaigns
will enable it to be measured more accurately
\citep{kra10,cal16}.

We emphasize in closing that the glitch rate law (\ref{eq:pin2})
may emerge from some other state-dependent Poisson process,
that has nothing to do with superfluid vortex avalanches and nuclear pinning,
such as starquakes
\citep{mid06,chu10b,gil21}.
\footnote{
Simulations of repeated, localized, crustal failure with a cellular automaton
indicate that starquakes are inconsistent with an inhomogeneous 
(but state-independent) Poisson process,
whose rate scales in proportion to the time-varying spin-down torque,
when events are counted over the spin-down time-scale;
see Appendix C in \citet{ker22}.
Given current glitch observations, however,
it is perfectly possible for starquakes to obey a state-dependent Poisson process,
either over the spin-down time-scale or, more conservatively, over $T_{\rm obs}$
\citep{ker22}.
}
If so, the Bayesian estimates of $\lambda_0$ and $X_{\rm cr}$ stand,
but the nuclear physics interpretation changes completely
and does not involve $f_{\rm p}$ and $E_{\rm a}$.
In the starquake picture, for example,
$X_{\rm cr}$ is likely to be proportional to the breaking strain of the crust.
Alternatively, two or more processes may coexist in the same or different pulsars.
The bimodal distribution of glitch sizes is one possible indicator that this occurs,
with smaller events associated arguably with starquakes,
and larger events associated with superfluid vortex avalanches
\citep{ash17,fue17,cel20,ant22,zho22,aru23}.
If so, 
the vortex unpinning contribution to $\gamma$ and hence $\lambda_0$ 
is overestimated by the Bayesian analysis in this paper,
which ascribes all glitches to vortex avalanches.
Even if superfluid vortex avalanches are the chief cause of glitches,
the physical picture underlying (\ref{eq:pin2}) leaves out several important and uncertain features,
such as pinning to magnetic flux tubes in the type II superconducting protons in the star's interior
\citep{sri90,rud98,dru18}
and general relativistic effects
\citep{ant18b}.
In order to disentangle these effects and many others,
more data are needed.

\acknowledgments
We thank the anonymous referee for insightful feedback
and for proposing the interesting analyses in \S\ref{sec:pin4c} and Appendix \ref{sec:pinappb}.
This research was supported by the Australian Research Council
Centre of Excellence for Gravitational Wave Discovery (OzGrav),
grant number CE170100004.

\bibliographystyle{mn2e}
\bibliography{glitchstat}

\appendix
\section{Glitch rate dispersion independent of age
 \label{sec:pinappa}}
Pulsar-to-pulsar variations in internal properties that depend weakly on age,
such as mass and chemical composition,
are expected theoretically to
produce some dispersion in the glitch rate $\gamma$ around the central value
implied by (\ref{eq:pin2}) and (\ref{eq:pin1b}).
The dispersion is parametrized phenomenologically by $\sigma$ in (\ref{eq:pin1b}),
in the absence of a microscopic theory of vortex avalanches
calibrated against controlled laboratory experiments.
With an eye to the moderate volume of data available,
we adopt a conservative strategy in \S\ref{sec:pin3}
and analyze the model (\ref{eq:pin2})--(\ref{eq:pin1c})
in the regime $\sigma \ll \gamma$,
where $\sigma$ drops out of (\ref{eq:pin1b}),
and $\lambda_{\rm ref}$, $a$, and $X_{\rm cr}$ are estimated.
However, $\sigma$ is an important quantity physically.
In this appendix, for the sake of completeness,
we estimate it by Bayesian inference using existing data,
while emphasizing that the analysis is preliminary and must be redone,
when more data become available.

The analysis follows the same procedure as in \S\ref{sec:pin3a},
except that $\sigma$ is left free in (\ref{eq:pin1b}) as a parameter to be estimated.
We analyze the same $N_{\rm g} \geq 1$ subsample as in \S\ref{sec:pin3a},
with the three quasiperiodic objects included, to facilitate comparison.
We apply the standard Jeffreys prior $p(\sigma) \propto \sigma^{-1}$
on the domain $10^{-14} \leq \sigma/(1 \, {\rm s^{-1}}) \leq 1$.
We confirm that the posterior PDF does not change,
when we widen the $\sigma$ domain.

\begin{figure}[ht]
\begin{center}
\includegraphics[width=16cm,angle=0]{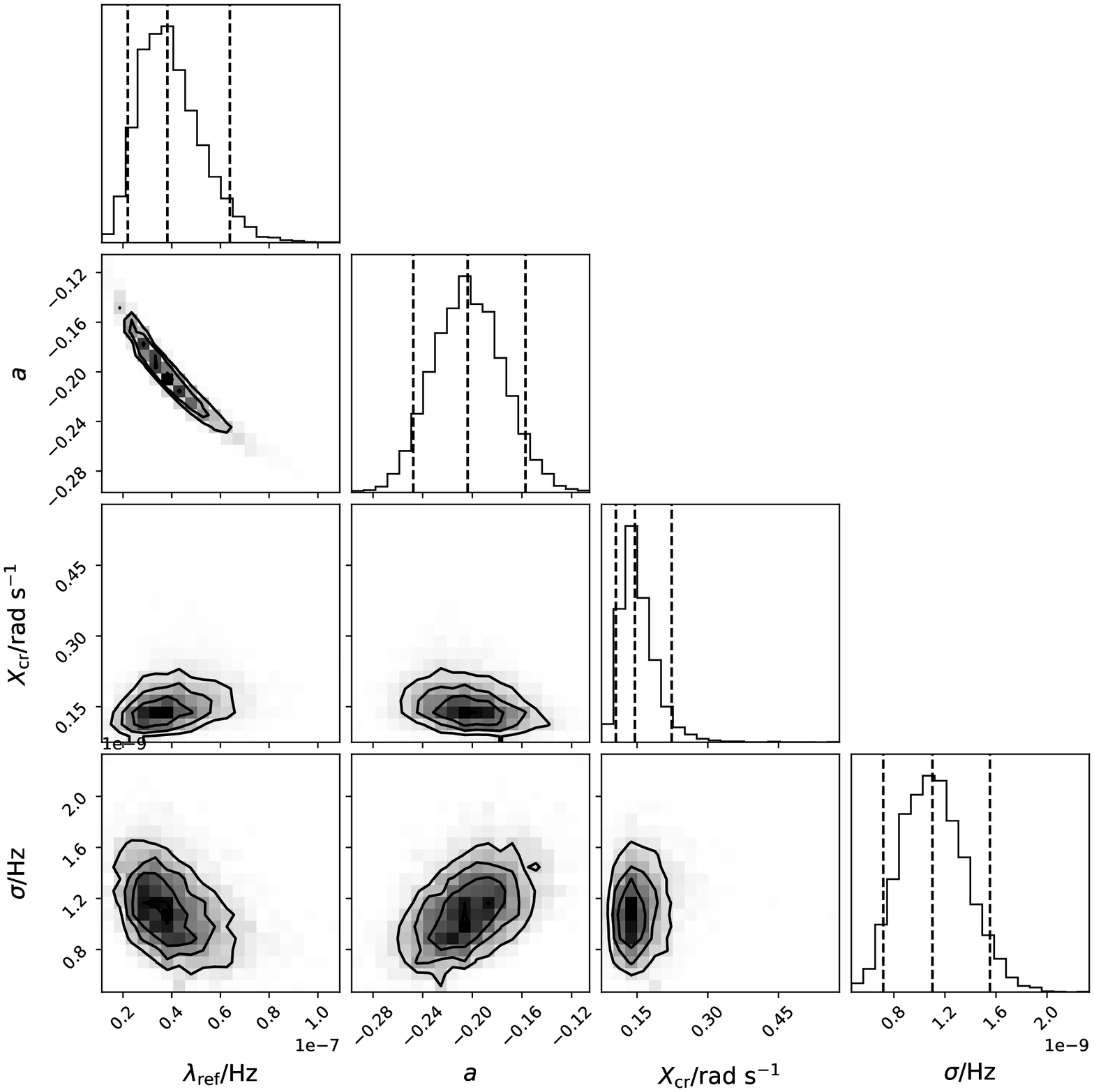}
\end{center}
\caption{
Posterior PDF of the phenomenological parameters
$\lambda_{\rm ref}$ (units: ${\rm s^{-1}}$), 
$a$ (dimensionless), 
$X_{\rm cr}$ (units: ${\rm rad \, s^{-1}}$), 
and $\sigma$ (units: ${\rm s^{-1}}$)
in the state-dependent Poisson model
for pulsars with $N_{\rm g} \geq 1$.
The layout matches Figure \ref{fig:pin2} with $\sigma$ added.
}
\label{fig:pin5}
\end{figure}

Figure \ref{fig:pin5} displays the posterior PDF for 
$\lambda_{\rm ref}$, $a$, $X_{\rm cr}$, and $\sigma$
as a traditional corner plot,
whose format matches Figure \ref{fig:pin2}, augmented to include $\sigma$.
The PDF is unimodal.
There are no strong covariances,
except for the familiar trade-off between $\lambda_{\rm ref}$ and $a$
noted in \S\ref{sec:pin3a}.
The PDF peaks at
$\lambda_{\rm ref} = 3.8^{+2.6}_{-1.6} \times 10^{-8} \, {\rm s^{-1}}$,
$a = -0.20^{+0.04}_{-0.05}$,
$X_{\rm cr} = 0.15^{+0.07}_{-0.05} \, {\rm rad \, s^{-1}}$,
and 
$\sigma = 1.1^{+0.5}_{-0.4} \times 10^{-9} \, {\rm s^{-1}}$.
Encouragingly, the peak of the PDF does not shift dramatically
within the subspace plotted in Figure \ref{fig:pin2};
the medians of $\lambda_{\rm ref}$, $a$, and $X_{\rm cr}$ from \S\ref{sec:pin3a}
are multiplied here by factors
$\approx 0.5$, $\approx 0.8$, and $\approx 0.9$ respectively.

Three tentative lessons can be drawn from the above exercise,
subject to the important caveats in the first paragraph.
(i) 
The estimate of $\sigma$ implies,
that much of the pulsar-to-pulsar variation in $\gamma$ arises from factors
that depend strongly on age, including possibly temperature,
rather than factors that depend weakly on age,
like mass and chemical composition.
Equations (\ref{eq:pin2}) and (\ref{eq:pin1c}) together with Figure \ref{fig:pin5} imply
$10^{-9} \lesssim \gamma / (1 \, {\rm s^{-1}}) \lesssim 10^{-8}$
for 
$10^3 \leq \tau / (1\, {\rm yr}) \leq 10^6$.
The inferred range of $\gamma$ is to be compared with the central estimate of $\sigma$,
viz.\ $\sigma \sim 10^{-9} \, {\rm s^{-1}}$.
There is preliminary evidence to support $\sigma \lesssim \gamma$ across most of
the pulsar population (i.e.\ $\tau \geq 10^3 \, {\rm yr}$).
(ii) 
If confirmed by future data,
the finding $\sigma \lesssim \gamma$ supports the assumption underpinning
the analysis excluding $\sigma$ in \S\ref{sec:pin3}.
(iii) 
We are approaching an era,
when enough data will be collected to estimate $\sigma$ with confidence,
within the context of the idealized model (\ref{eq:pin2})--(\ref{eq:pin1c}).
Expanding the glitch catalog through a new generation of radio timing campaigns
will improve the accuracy of the $\sigma$ estimate appreciably
\citep{kra10,cal16}.

\section{Giant glitches
 \label{sec:pinappb}}
Statistical studies of the aggregate glitch population reveal that the 
event size PDF is bimodal
\citep{esp11,ash17,fue17,eya19,cel20,ant22,zho22,aru23}.
By fitting a two-component, skewed, Gaussian mixture model to 
the aggregate data,
one finds that $\approx70\%$ of glitches have absolute size $\Delta\Omega_{\rm c}$ satisfying
$\Delta\Omega_{\rm c} / (2\pi \, {\rm Hz}) \leq 10^{-5}$
and are termed `normal',
while $\approx 30\%$ satisfy
$\Delta\Omega_{\rm c} / (2\pi \, {\rm Hz}) \geq 10^{-5}$
and are termed `giant'.
\footnote{
The two components of the skewed, Gaussian mixture model have means of
$\langle \Delta\Omega_{\rm c} \rangle / (2\pi \, {\rm Hz}) = 4.1\times 10^{-9}$
and 
$\langle \Delta\Omega_{\rm c} \rangle / (2\pi \, {\rm Hz}) = 3.9\times 10^{-5}$;
see Table 1 and Figure 1 of \citet{ash17}.
An extreme deconvolution analysis yields similar, bimodal results
for the PDF of $\Delta\Omega_{\rm c}/\Omega_{\rm c}$
\citep{aru23}.
}
Sometimes the giant glitches are also termed `Vela-like',
because they occur in pulsars whose characteristic spin-down ages
and inferred magnetic dipole moments resemble those of Vela
\citep{ash17}.
It is tempting to speculate that normal and giant glitches are caused
by different physical mechanisms,
although the same mechanism can produce both, 
e.g.\ superfluid vortex avalanches combined with nonlinear mututal friction 
\citep{cel20}.
Either way, the question arises:
do the parameter estimates in \S\ref{sec:pin3} change,
when giant glitches are excluded?
We explore this question briefly in this appendix.

Giant glitches are not defined uniquely.
For example,
one may impose a threshold on absolute size $\Delta\Omega_{\rm c}$
\citep{esp11,ash17,fue17,cel20}
or fractional size $\Delta\Omega_{\rm c} / \Omega_{\rm c}$
\citep{eya19}.
The former option is appropriate,
if giant glitches are associated with their own, separate mechanism
triggered above a specific, physical scale,
e.g.\ a hydrodynamic instability \citep{and03,gla09}.
The latter option is appropriate,
if giant glitches are the large-system limit of the normal, scale-invariant glitch mechanism,
e.g.\ system-spanning vortex avalanches
\citep{war11}.
One may also categorize giant glitches per event
(individual $\Delta\Omega_{\rm c}$ exceeds a threshold)
or per pulsar
(average $\langle \Delta\Omega_{\rm c} \rangle$ exceeds a threshold).
The categorization schemes are different;
not all Vela glitches are giant, for example
\citep{esp11,ash17,fue17,eya19,cel20}.
\citet{fue17} counted 70 large glitches in 38 objects,
whereas \citet{ash17} counted $\sim 150$ large glitches in $\sim 50$ objects
with a related but different definition.

The analysis in \S\ref{sec:pin2} and \S\ref{sec:pin3} infers glitch rates from waiting times,
i.e.\ multiple events per pulsar.
We are therefore obliged to categorize giant glitches on a per-pulsar basis
not a per-event basis.
We define pulsars that exhibit giant glitches as being those objects,
whose average absolute glitch size satisfies 
$\langle \Delta\Omega_{\rm c} \rangle / (2\pi) \geq 10^{-5} \, {\rm Hz}$.
The definition is arbitrary, of course, but it is reasonable and conforms with
similar definitions in the literature cited above.
It yields 41 pulsars with a total of 163 events.
The sample contains two quasiperiodic objects
(PSR J0537$-$6910 and PSR J0835$-$4510),
the 13 objects listed in Table \ref{tab:pin1},
which are Vela-like in a sense to be specified below,
and 26 other objects (of which 20 have $N_{\rm g}=1$).

\begin{table}
\begin{center}
\begin{tabular}{lcccc} \hline
 PSR J & $N_{\rm g}$ & $\langle \Delta \Omega_{\rm c} \rangle / (2\pi)$  ($10^{-5} \, {\rm Hz}$) 
  & Age (kyr) & Magnetic field ($10^{12} \, {\rm G}$) \\ \hline
 1016$-$5857 & 2 & 1.2 & 21 & 3.0 \\
 1048$-$5832 & 6 & 1.5 & 20 & 3.5 \\
 1301$-$6305 & 2 & 1.6 & 11 & 7.1 \\
 1357$-$6429 & 2 & 1.6 & 7.3 & 7.8 \\
 1420$-$6048 & 5 & 2.1 & 13 & 2.4 \\
 1614$-$5048 & 2 & 2.5 & 7.4 & 11 \\
 1709$-$4429 & 5 & 1.7 & 18 & 3.1 \\
 1730$-$3350 & 4 & 1.5 & 26 & 3.5 \\
 1801$-$2451 & 5 & 1.7 & 16 & 4.0 \\
 1803$-$2137 & 5 & 2.6 & 16 & 4.3 \\
 1826$-$1334 & 6 & 2.0 & 21 & 2.8 \\
 1932$+$2220 & 3 & 2.2 & 40 & 2.9 \\
 2021$+$3651 & 4 & 1.9 & 17 & 3.2 \\
 \hline
\end{tabular}
\end{center}
\caption{
Representative sample of Vela-like pulsars that exhibit giant glitches with average size
$\langle \Delta\Omega_{\rm c} \rangle / (2\pi) \geq 10^{-5} \, {\rm Hz}$
and have characteristic spin-down ages and magnetic dipole moments
resembling those of Vela.
The ages (in units of ${\rm kyr}$) and surface magnetic field strengths
(in units of $10^{12} \, {\rm G}$)
are drawn from the ATNF Pulsar Catalogue \citep{man05}.
}
\label{tab:pin1}
\end{table}

\begin{figure}[ht]
\begin{center}
\includegraphics[width=16cm,angle=0]{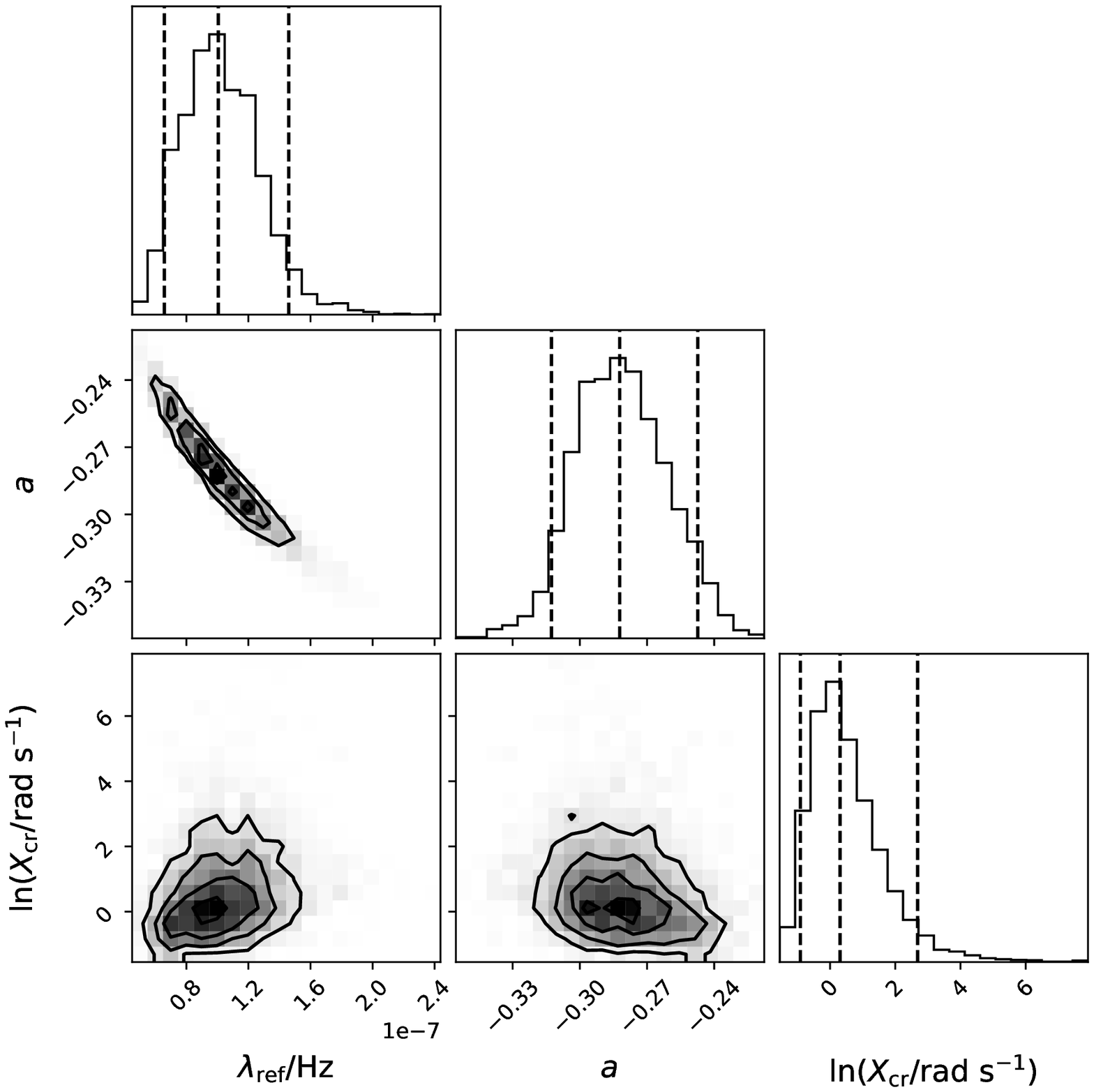}
\end{center}
\caption{
Posterior PDF of the phenomenological parameters
$\lambda_{\rm ref}$ (units: ${\rm s^{-1}}$), 
$a$ (dimensionless), 
and
$X_{\rm cr}$ (units: ${\rm rad \, s^{-1}}$)
in the state-dependent Poisson model
for pulsars with $N_{\rm g} \geq 1$,
assuming $\sigma \ll \gamma$
but excluding the 41 objects with giant glitch activity satisfying
$\langle \Delta\Omega_{\rm c} \rangle / (2\pi) \geq 10^{-5} \, {\rm Hz}$.
The layout matches Figure \ref{fig:pin2},
except that the $X_{\rm cr}$ scale is logarithmic (base $e$).
}
\label{fig:pin6}
\end{figure}

The 163 events in the giant glitch sample are too few to analyze in isolation
via the Bayesian recipe in \S\ref{sec:pin2}.
Instead we adopt the same approach as for the objects with 
quasiperiodic glitch activity in \S\ref{sec:pin3c}:
we exclude the 41 objects from the $N_{\rm g} \geq 1$ sample analyzed in \S\ref{sec:pin3a}
and ask whether or not the inferred parameters change significantly.
The results are presented as a traditional corner plot in Figure \ref{fig:pin6}.
The PDF is unimodal and resembles Figure 2 qualitatively,
with 
$\lambda_{\rm ref} = 1.0^{+0.45}_{-0.34} \times 10^{-7} \, {\rm s^{-1}}$,
$a=-0.28^{+0.03}_{-0.03}$,
and $X_{\rm cr} = 1.4^{+13}_{-1.0} \, {\rm rad\, s^{-1}}$
(median and 90\% confidence interval).
Overall, the results are consistent with \S\ref{sec:pin3}.
The parameter estimates overlap with the error bars in \S\ref{sec:pin3c}, 
where all three quasiperiodic objects --- but no others --- are excluded.
This is expected, because the two quasiperiodic objects that also exhibit
giant glitch activity account for 65 out of the 163 events excluded from Figure \ref{fig:pin6}.
The $\lambda_{\rm ref}$ and $a$ estimates also overlap with the error bars 
for the full $N_{\rm g} \geq 1$ sample in \S\ref{sec:pin3a},
whereas the median $X_{\rm cr}$ is $\approx 9$ times higher.
Again this is expected.
Giant glitches occur in objects with relatively high $|\dot{\Omega}_{\rm c}|$
[see Figure 3 in \citet{ash17}, for example],
which contribute meaningfully to the second term in (\ref{eq:pin2}),
just like the quasiperiodic objects in \S\ref{sec:pin3c}.
More data are needed to test, if a statistically significant tail of pulsars with
relatively high $|\dot{\Omega}_{\rm c}|$ emerges in the future,
which keeps $X_{\rm cr} \sim 0.1 \, {\rm rad \, s^{-1}}$ relatively low,
as in \S\ref{sec:pin3a}, 
even after giant glitches are excluded.

Not all of the 41 objects excluded from Figure \ref{fig:pin6} have 
characteristic spin-down ages and magnetic dipole moments resembling those of Vela.
As a second check, therefore, we repeat the baseline $N_{\rm g} \geq 1$ analysis in \S\ref{sec:pin3a}
while excluding only the 13 objects listed in Table \ref{tab:pin1} with a total of 51 events.
The subsample in Table \ref{tab:pin1} is categorized as Vela-like in line with the literature
\citep{ash17},
because the average glitch size per member qualifies as giant,
with 
$\langle \Delta\Omega_{\rm c} \rangle / (2\pi) \geq 10^{-5} \, {\rm Hz}$;
the characteristic spin-down age lies between $7.3 \, {\rm kyr}$ and $40 \, {\rm kyr}$
(cf.\ $11 \, {\rm kyr}$ for Vela);
and the surface dipole magnetic field strength lies between 
$2.4 \times 10^{12} \, {\rm G}$ and $1.1\times 10^{13} \, {\rm G}$
(cf.\ $3.4 \times 10^{12} \, {\rm G}$ for Vela).
The criteria are not unique, of course, but they are representative.
Again, the results are consistent with \S\ref{sec:pin3a}.
The PDF (not plotted for brevity) is unimodal and resembles Figure 2 qualitatively,
with 
$\lambda_{\rm ref} = 9.3^{+4.5}_{-3.3} \times 10^{-8} \, {\rm s^{-1}}$,
$a=-0.28^{+0.04}_{-0.03}$,
and $X_{\rm cr} = 0.17^{+0.11}_{-0.06} \, {\rm rad\, s^{-1}}$
(median and 90\% confidence interval).
All three parameter estimates overlap with the error bars for the full
$N_{\rm g} \geq 1$ sample in \S\ref{sec:pin3a},
which is expected, as Table \ref{tab:pin1} excludes 51 out of a total of 541 events.
The subsample is too small to be analyzed in isolation via the Bayesian recipe in \S\ref{sec:pin2}.
Four out of the 13 objects in Table \ref{tab:pin1} have $N_{\rm g}=2$,
and all 13 objects have $N_{\rm g} \leq 6$,
so it is impossible to draw reliable conclusions about a shared, Vela-like propensity
for quasiperiodic glitch activity.

\end{document}